\documentclass[review,authoryear,12pt]{elsarticle}
\usepackage{amssymb}
\usepackage{amsmath}
\usepackage{bm}
\usepackage{graphicx}
\usepackage{subfigure}
\usepackage{longtable}
\usepackage{setspace}
\usepackage{natbib}
\usepackage{lineno}
\usepackage{setspace}
\usepackage{natbib}

\DeclareMathOperator*{\argmin}{arg\,min}

\journal{Atmospheric Environment}

\begin{document}

\begin{frontmatter}

%%\relsize{0.95}
%\bibliographystyle{plain}
%\title{Conditional Simulations of Anisotropic Spatial Data with Global Gradient-Curvature Constraints: Application to Data Gap Filling in Satellite Imaginary}
%\author{Milan \v{Z}ukovi\v{c}\textsuperscript{1},
%Dionissios T. Hristopulos\textsuperscript{2}}
%\maketitle
%\begin{center}
%\small{
%\textsuperscript{1}Geostatistics Research Unit, Department of Mineral Resources Engineering,
%Technical University of Crete\\
%mzukovic@mred.tuc.gr
%
%\textsuperscript{2}Geostatistics Research Unit, Department of Mineral Resources Engineering,
%Technical University of Crete\\
%dionisi@mred.tuc.gr}
%\end{center}

\title{A Directional Gradient-Curvature Method for Gap Filling of
 Gridded Environmental Spatial Data with Potentially Anisotropic Correlations}
\author[UPJS,SORS]{Milan \v{Z}ukovi\v{c}}
\ead{milan.zukovic@upjs.sk}
\author[TUC]{Dionissios T. Hristopulos\corref{cor}}
\ead{dionisi@mred.tuc.gr}
\ead[url]{http://www.mred.tuc.gr/home/hristopoulos/dionisi.htm}
\address[UPJS]{Institute of Physics, Faculty of Science,
 Pavol Jozef \v{S}af\'{a}rik University, Park Angelinum 9, 040 01 Ko\v{s}ice, Slovakia}
\address[SORS]{SORS Research a.s., Moyzesova 38, 040 01 Ko\v{s}ice, Slovakia}
\address[TUC]{Geostatistics Research Unit, Department of Mineral
Resources Engineering,  Technical University of Crete, Chania 73100,
Greece} 

%\tnotetext[current]{Current Address: Institute of Physics,
%Faculty of Science,
% Pavol Jozef \v{S}af\'{a}rik
%University, Park Angelinum 9, 040 01 Ko\v{s}ice, Slovak Republic}
\cortext[cor]{Corresponding author.}

\begin{abstract}

We introduce the Directional Gradient-Curvature (DGC) method, a
novel approach for filling gaps in gridded environmental data. DGC is
based on an objective function that measures the distance between
the directionally segregated normalized squared gradient and
curvature energies of the sample and entire domain data. DGC employs
data-conditioned simulations, which sample the local minima
configuration space of the objective function instead of the full
conditional probability density function. Anisotropy and
non-stationarity can be captured by the local constraints  and the
direction-dependent global constraints. DGC is computationally
efficient and requires minimal user input, making it suitable for
automated processing of large (e.g., remotely sensed) spatial data
sets. Various effects are investigated on synthetic data. The
gap-filling performance  of DGC is assessed in comparison with established
classification and interpolation methods using synthetic
and real satellite data, including a skewed distribution of daily column
ozone values. It is shown that DGC is competitive in terms of cross validation 
performance.

\end{abstract}


\end{frontmatter}

\linenumbers

\section{Introduction}

Atmospheric data, whether they are obtained by means of ground or remote sensing 
 methods, often include data gaps.  Such gaps arise due to
different reasons, e.g. incomplete time series, spatial
irregularities of sampling pattern, equipment limitations or sensor
malfunctions~\citep*{Jun04,Lehman04,Albert12,Bechle13}. For example, remote sensing images may include obscured areas 
due to cloud cover, whereas gaps also appear between satellite paths where there
is no coverage for a specific period~\citep*{Emili11}. 
The impact of missing data on the estimate of statistical averages and trends can be significant~\citep*{Sickles07}. 
There is an interest in the development of new methods for filling gaps in 
atmospheric data and their comparison with existing imputation  methods~\citep*{Junninen04}. 
Particularly for frequently
collected, massive remotely sensed data, the efficient filling of
the gaps is a challenging task. Traditional geostatistical
interpolation methods such as kriging, e.g.~\citep*{wack03}, can be
impractical due to high computational complexity, restriction to
Gaussian data, as well as various subjective choices in variogram
modeling and interpolation search radius~\citep*{dig07}. In
particular, computationally efficient methods are needed for filling
gaps in very large data sets~\citep*{cressie08,hartman08}.

In the following, we consider   a set of sampling points
$G_{s}=\{\vec{s}_{i}, \, i=1,\ldots, N\} $, where $\vec{s}_{i}=(x_i,
y_i) \in {\mathbb{R}}^2$. The points are scattered on a rectangular
grid $\tilde{G}$ of size $N_{G}=L_{x} \times L_{y}$, where $L_{x}$
and $L_{y}$ are respectively the horizontal and vertical dimensions
of the rectangle (in terms of the unit length), such that $N_{G}>N$.
Let $G_{p}=\{\vec{s}_{p}, p=1,\ldots,P\} $ be the set of prediction
points, representing locations of missing values, such that
$\tilde{G}=G_{s} \cup G_{p}$. The data, ${\bf Z}(G_{s})=\{z_{i},
\forall \vec{s}_{i} \in G_{s} \},$ are considered as a realization
of the continuous random field $Z({\vec{s}_i})$. To reduce the
dimensionality of the configuration space, we discretize the
continuously valued field. For applications that do not require high
resolution, e.g. environmental monitoring and risk management, ${\bf
Z}(G_{s})$ can be discretized into a small (e.g., eight) number
$N_c$ of levels (classes). Continuous distributions are obtained at
the limit $N_c \rightarrow \infty$. In the current study, the
spatial prediction of
 missing values is posed as a spatial classification problem
for ranked numerical data. Continuous interpolation is approximated
by considering an arbitrarily high number of levels.

The discretization classes ${\mathcal C}_q, \, q=1,\ldots,N_c$
correspond to the intervals ${\mathcal C}_q=(t_{q}, t_{q+1}]$ for
$q=2,\ldots,N_c-1$, ${\mathcal C}_1=(-\infty, t_{2} ],$ and
 ${\mathcal C}_{N_c}=(t_{N_{c}}, \infty).$ The classes are
 defined with respect to threshold levels
$t_k, \, k=2,\ldots,N_{c}$. All the classes have a uniform width
 except  ${\mathcal C}_1$ and ${\mathcal
C}_{N_c}$ which extend to negative and positive infinity
respectively, to include values outside the observed interval
$[z_{\min}, z_{\max}]$. More general class definitions can be
investigated. The \textit{class identity field} $I(\vec{s})$ takes
integer values $q=1,\ldots,N_c$ that represent the respective class
index. In particular, $I(\vec{s_{i}})= q$ implies that $z_{i} \in
{\mathcal C}_q.$ The prediction problem is equivalent to assigning a
class label at each point in $G_{p}$. A map of the process $Z$ can
be generated consisting of equivalent-class (isolevel) contours.

\section{The Directional Gradient-Curvature Model}
 The Directional Gradient-Curvature (DGC) model is
inspired by Spartan spatial random fields (SSRF)~\citep*{dth03},
which are based on short-range interactions between the field
values. The SSRF model is parametric and represents stationary,
continuous and isotropic Gaussian random fields. To relax these
assumptions, we introduce an almost non-parametric approach that
aims at matching short-range correlations in $G_{s}$ with those of
$\tilde{G}$. This idea was recently successfully applied to spatial 
random fields still assuming spatial isotropy~\citep*{mz-dth12}.
The present model extends this approach by relaxing even the isotropic 
assumption through incorporating anisotropic dependence. In particular, the
correlations used in DGC represent the normalized squared gradient
and curvature energies of the discretized class identity field along
different directions. Let $a_{n}$ be the lattice step in direction
$\vec{e}_{n}$. The local square gradient and curvature terms in each
of the $d$ directions used are given by

\begin{equation}
\label{VN-G} G_{n}(I;\vec{s}_{i})  =  \frac{\left[ I(\vec{s_i} +
a_{n}\vec{e}_{n}) - I(\vec{s_i})  \right]^{2}}{a_{n}^{2}}, \quad
n=1,\ldots,d
\end{equation}

\begin{equation}
\label{VN-C} C_{n}(I;\vec{s}_{i})  =   \frac{\left[I (\vec{s_i} +
a_{n}\vec{e}_{n}) + I(\vec{s_i}-a_{n}\vec{e}_{n}) -2
I(\vec{s_i})\right]^{2}}{a_{n}^{4}}, \quad n=1,\ldots,d.
\end{equation}

The normalized squared gradient, $\overline{G}_{n}({\bf I}_{g}),$
and curvature, $\overline{C}_{n}({\bf I}_{g})$ energies in a
direction $\vec{e}_{n}$, are defined as averages of the above over
the grid $\tilde{G}$. ${\bf I}_{g}=[I(\vec{s}_{1}) \ldots
I(\vec{s}_{N_{G}})]\, \forall \, \vec{s}_{i} \in \tilde{G},$
represents the values of the class identity field on the entire
grid, ${\bf I}_{p}=[I(\vec{s}_{1}) \ldots I(\vec{s}_{P})]\, \forall
\, \vec{s}_{i} \in G_{p},$ is the class field at the prediction
sites, and ${\bf I}_{s}=[I(\vec{s}_{1}) \ldots I(\vec{s}_{N})]\,
\forall \, \vec{s}_{i} \in G_s$ are the class identity values at the
sampling sites. The matching of the gradient and curvature
constraints on $G_s$ and $\tilde{G}$ is based on  the following
\textit{objective functional}:

%\begin{equation}
%\label{cost} U({\bf I}_{p}|{\bf I}_{s}) = \sum\limits_{n=1}^{d} w_1
%\left[ 1-\frac{\overline{G}_{n}({\bf I}_{g})}{\overline{G}_{n}({\bf
%I}_{s})}\right]^2 + w_2 \left[ 1-\frac{\overline{C}_{n}({\bf
%I}_{g})}{\overline{C}_{n}({\bf I}_{s})}\right]^2,
%\end{equation}

\begin{align}
\label{cost} U({\bf I}_{p}|{\bf I}_{s}) & = \sum\limits_{n=1}^{d} \,
\left[ w_1 \, \phi \big(\overline{G}_{n}({\bf I}_{g}),
\overline{G}_{n}({\bf I}_{s})\big) + w_2 \, \phi
\big(\overline{C}_{n}({\bf I}_{g}),
\overline{C}_{n}({\bf I}_{s})\big) \right],\\
\label{eq:phi} \phi(x,x') & = \left\{ \begin{array}{ll} \left(
1-x/x'\right)^{2}, & x'\neq 0 \\
 x^{2}, & x'=0. \end{array} \right.
\end{align}

In the above, $w=(w_1,w_2)$ represent gradient and curvature
 weights ($w_1,w_2 \ge 0$, $w_1+w_2=1$), and $d$ is the number of the
 directions used. We use~\eqref{eq:phi} to measure the deviation
 between the $G_{s}-$ and $\tilde{G}-$based values instead of
 $\phi=(x-x')^{2}$, because the gradient and curvature
 can have very different magnitudes, depending on the units used;
 this means that one term may dominate in the optimization.
  By using normalized constraints we compensate for the possible disparity of
  magnitudes between gradient and curvature. If one of the sample quantities is zero, the second line
  of~\eqref{eq:phi} is used to avoid a singular denominator.

 We select $w=(0.5,0.5)$ and $d=4$, representing four
directions with the following angles (with respect to the positive
x-axis): $0^{\circ}$, $45^{\circ}$, $90^{\circ}$, and $135^{\circ}$.
Given the above, the classification problem  is equivalent to
determining the optimal configuration $\hat{{\bf I}}_{p}$ that
corresponds to the minimum of~(\ref{cost}):
%$\hat{I}_{p} =\argmin_{{\bf I}_{p}}U({\bf I}_{p}|{\bf I}_{s}).$

\begin{equation}
\label{optim} \hat{{\bf I}}_{p} = \argmin_{{\bf I}_{p}}U({\bf
I}_{p}|{\bf I}_{s}).
\end{equation}

The optimization of~(\ref{cost}) is conducted numerically. The
choice of the initial configuration is important to obtain a
reliable, fast and automatic algorithm: it should prevent the
optimization from getting trapped in poor local minima and minimize
the relaxation path in configuration space to the equilibrium, and
it should also minimize the need for user intervention. The sampling
points retain their values ${\bf I}_{s}$. Assuming a certain degree
of spatial continuity, common in geospatial data sets, the initial
state of ${\bf I}_{p}$ is determined based on the sample states in
the immediate neighborhood of the individual prediction points. The
neighborhood of ${\vec{s}_p}$ is determined by an $m \times m$
stencil ($m=2l+1$) centered at ${\vec{s}_p}$. Then, the initial
value at a prediction point is assigned by majority rule, based on
the prevailing value of its sample neighbors
 inside the stencil. The stencil size
is chosen automatically, reflecting the local sampling density and
the  distribution of \textit{class identity} values. Namely, it is
adaptively set to the smallest size that contains a finite number of
sampling points with a prevailing value. If no majority is reached
up to some neighborhood size $m_{\max}\times m_{\max}$, the initial
value is assigned (i) randomly from the range of the labels with tie
votes or (ii) from the entire range of labels $1,...,N_c$, if
majority is not reached due to absence of sampling points within the
maximum stencil. The above method of initial state assignment will
be referred to as majority rule with adaptable stencil size (MRASS).

The updating of class identity states on $G_{p}$ uses the ``greedy''
Monte Carlo (MC) method~\citep*{papa82}, which unconditionally
accepts a new state if the latter lowers the cost function. The
greedy MC algorithm may cause the termination of the DGC algorithm
at local minima of the objective function~(\ref{cost}). Targeting
exclusively global minima (e.g. by simulated annealing) unduly
emphasizes  exact matching of the energies on the entire domain with
those in the sample domain; however, the latter are subject to
sampling fluctuations and measurement errors.

The algorithm performs a random walk through the grid $G_{p}$. It
terminates if $P$ consecutive update trials do not produce a single
successful update. If the computational budget is a concern, the
algorithm can terminate when a pre-specified maximum number of Monte
Carlo steps is exceeded. In either case, the generated realization
is accepted only if the residual value of the cost function is below
a user-defined tolerance level $tol$. Otherwise, the realization is
rejected and a new one is generated. The algorithm generates $M$
different realizations. The median values from all the accepted
realizations at each missing-value point represent the prediction of
the algorithm. Associated confidence intervals are also derived.

The main steps of the procedure  described above are summarized by
means of the following algorithm.
\begin{enumerate}

\item  Define the number of realizations $M$, the number of
classes $N_c$, the maximum stencil size $m_{\max}$, the residual
cost function tolerance $tol$ and
the maximum number of Monte Carlo steps $i_{\max}$ (optional).
\item Discretize ${\bf Z}(G_{s})$
to obtain the sample \textit{class identity field} ${\bf I}_{s}.$
\item Calculate the directional sample energies $\overline{G}_{n}({\bf I}_{s})$,
$\overline{C}_{n}({\bf I}_{s})$, $n=1,\ldots,d.$\footnote{
The algorithm checks if the number of samples for calculating
$\overline{G}_{n}({\bf I}_{s})$, $\overline{C}_{n}({\bf I}_{s})$ is
sufficient for obtaining reliable estimates.}
\item  Initialize the simulated realization index $j =1$.
\item {\bf while} $j \leq M$ repeat the following steps:
    \begin{enumerate}
                        \item Assign initial values $\hat{\bf I}_{p}^{(0)}$
                        to the prediction points in $G_{p}$  based on MRASS.
                        \item Calculate the initial energy values
                        $\overline{G}_{n}({\bf I}^{(0)}_g)$, $\overline{C}_{n}({\bf
                        I}^{(0)}_g)$, $n=1,\ldots,d$,
                        and the objective function $U^{(0)}=U(\hat{\bf I}^{(0)}_{p}|{\bf I}_{s}).$
                        \item  Initialize the simulated state index $i =0,$
                        and the rejected states index $i_r =0$.
                        \item {\bf while} $(i_r<P) \wedge (i \leq i_{\max})$ repeat the following updating steps:
                            \begin{enumerate}
                                    \item Generate a new state $\hat{\bf I}_{p}^{(i+1)}$
                                    by randomly, i.e., with probability 0.5, adding $\pm 1$ to
                                    the state $\hat{{\bf I}}_{p}^{(i)}$, maintaining the condition
                                    $1\leq \hat{I}^{(i+1)}(\vec{s}_{j})\leq N_{c}, \;
                                    \forall \vec{s}_{j} \in G_{p}.$
%        \item Generate a new state
%            $\hat{I}^{(i+1)}_{p}$  by randomly perturbing $\hat{I}^{(i)}_{p}.$
                                        \item Calculate
                                    $\overline{G}_{n}({\bf I}^{(i+1)}_g)$,
                                    $\overline{C}_{n}({\bf I}^{(i+1)}_g)$,
                                    $n=1,\ldots,d.$
                                    \item Calculate $U^{(i+1)}=U(\hat{{\bf I}}^{(i+1)}_{p}|{\bf I}_{s}).$
                                        \item {\bf If} $U^{(i+1)} < U^{(i)}$ accept the
                                        new state $\hat{{\bf I}}^{(i+1)}_{p}$; $i_r \rightarrow 0$; \newline
                                        {\bf else} $\hat{{\bf I}}^{(i+1)}_{p}=\hat{{\bf I}}^{(i)}_{p}$; $U^{(i+1)}=U^{(i)}$; $i_r \rightarrow i_r+1$; {\bf end}.
                                        \item $i \rightarrow i+1$;
                            \end{enumerate}
                                {\bf end while}
                    \item {\bf If} $U^{(i)} < tol$ store the realization $\hat{{\bf I}}^{*}_{p}(j)=\hat{{\bf I}}^{(i)}_{p}$; $j=j+1$; \newline
                                {\bf else} return to 5 (a); {\bf end}.
        \end{enumerate}
        {\bf end while}
\item Evaluate the statistics   from the realizations $\hat{{\bf I}}^{*}_{p}(j),\ j=1,...,M.$
\end{enumerate}

The DGC  method lies between interpolation and conditional
simulation. Interpolation methods provide a single optimal
configuration of the missing values, e.g.,  kriging is based on the
minimization of the mean square error. Conditional simulation based
on  Markov Chain Monte Carlo methods aims to sample the entire
configuration space and reconstruct the joint conditional
probability density function of the missing data. DGC on the other
hand samples the configuration space that corresponds to local
minima of the objective function. Since DGC returns multiple
realizations, we can characterize it as a stochastic method.
However, in DGC  the sampling of the configuration space is
restricted to the subspace of local minima. The afforded
dimensionality reduction is responsible for the computational
efficiency of the method.

\section{DGC validation methodology}
In this section we conduct numerical experiments, in which a portion
of the data is set aside to be used for validation of the
classification/interpolation algorithms tested. The performance of
DGC is evaluated by calculating the misclassification rate $F^{*} =
1/P\sum_{\vec{s}_{p} \in
G_{p}}\left[1-\delta\big(I(\vec{s}_p),\hat{I}(\vec{s}_p)\big)
\right]$, where $I(\vec{s}_p)$ is the true class identity value at
the validation points, $\hat{I}(\vec{s}_p)$ is the classification
estimate and $\delta(I,I')=1$ if $I=I'$, $\delta(I,I')=0$ if $I \neq
I'$. The gap-filling of DGC is compared with the $k$-nearest
neighbor (KNN) \citep*{dasa91} and fuzzy $k$-nearest neighbor (FKNN)
\citep*{kell85} classification algorithms. We chose the $k$  values
that minimize the cross validation errors to obtain the lowest
achievable errors by KNN and FKNN. The KNN and FKNN algorithms are
applied using the Matlab\circledR\, function
\verb+fknn+~\citep*{akbas}.

The interpolation performance is compared with the inverse distance
weighted (ID)~\citep*{shep68}, nearest neighbors (NN), bilinear
(BL), bicubic (BC), and biharmonic spline (BS)\citep*{sand87}
methods. For the Gaussian synthetic data we also include the
ordinary kriging (OK) method~\citep*{wack03}. Given the Gaussian
distribution and knowledge of the covariance parameters, OK provides
optimal predictions and thus also a standard for comparing DGC
estimates. The NN, BL, BC and BS interpolation algorithms were
implemented by means of the Matlab\circledR\, function
\verb+griddata+. For ID we used the Matlab\circledR\, function
\verb+fillnans+~\citep*{Howat}. Finally, for OK we used the routines
available in the Matlab\circledR\, library
\verb+vebyk+~\citep*{Sidler09}.

Let $\hat{Z}(\vec{s}_p)$ be the estimate
 of the continuous field calculated from the back transformation
\begin{equation}
 \hat{Z}(\vec{s}_p) = [t_{\hat{I}_{Z}(\vec{s}_p)}+t_{\hat{I}_{Z}(\vec{s}_p)+1}]/2,\ p=1,\ldots,P.
 \end{equation}
If $Z(\vec{s}_p)$ is the true value at  $\vec{s}_p$ the estimation
error is $\epsilon(\vec{s}_p)= Z(\vec{s}_p) - \hat{Z}(\vec{s}_p).$
For $N_c>>1$ we calculate the following prediction errors: average
absolute error 
\begin{equation}
{\rm AAE} = (1/P)\sum_{\vec{s}_{p} \in G_{p}}|\epsilon(\vec{s}_p)|,
\end{equation} 
average relative error
\begin{equation}
{\rm ARE}=(1/P)\sum_{\vec{s}_{p} \in G_{p}}\epsilon(\vec{s}_p)/Z(\vec{s}_p), 
\end{equation}
average absolute relative error
\begin{equation}
{\rm AARE} =(1/P)\sum_{\vec{s}_{p} \in G_{p}}|\epsilon(\vec{s}_p)|/Z(\vec{s}_p), 
\end{equation}
root average squared error 
\begin{equation}
{\rm RASE} =\sqrt{\sum_{\vec{s}_{p} \in G_{p}}(1/P)\,\epsilon^2(\vec{s}_p)}, 
\end{equation}
and linear correlation coefficient $R$.

If $S$ sample configurations are considered, the mean values of the
validation measures (i.e., the MAAE, MARE, MAARE, MRASE, and MR) are
calculated by averaging over the sample configurations. To focus on
the local performance of DGC, we use the respective ``local''
errors, i.e., MAE, MRE, MARE, and RMSE, in which the spatial average
is replaced by the mean  over predictions obtained from $M$
different simulations. Furthermore, we record the optimization CPU
time, $T_{cpu}$, and the number of Monte Carlo steps (MCS).

 The computations are performed in
Matlab\textregistered\ programming environment on a desktop computer
with 3.25 GB RAM and an Intel\textregistered Core\texttrademark 2
Quad CPU Q9650 processor with an 3 GHz clock.

\section{Results}
\label{results}

\subsection{Synthetic Data}
\label{synth_data}

DGC performance is first studied on synthetic data sampled on
regular grids. The data  are simulated from the Gaussian random
field $Z \sim N(m=50,\sigma=10)$ with Whittle-Mat\'{e}rn covariance
given by $c_{\rm Z} (\vec{r})=\sigma^{2} \,
\frac{{2}^{1-\nu}}{\Gamma(\nu)}\, h^{\nu}\, K_{\nu}(h)$, where
$h=\sqrt{r^{2}_{1}/\xi_{1}^{2} + r^{2}_{2}/\xi_{2}^{2}}$ and
$\vec{r}=(r_{1},r_{2})$ is the lag  distance between two points.
$K_{\nu}$ is the modified Bessel function of the second kind and of
order $\nu$, where  $\nu=2.5$ is the covariance smoothness
parameter. The principal axes of anisotropy are aligned with the
coordinate axes. The correlation length in the vertical direction is
set to $\xi_{2}=2$ and in the horizontal direction to $\xi_{1}=4.$
The field is sampled on a square grid $\tilde{G}$, with $N_{G}=50
\times 50$ nodes using the spectral method~\citep*{drum87}. Missing
data samples ${\bf Z}(G_{s})$ of size $N=N_{G} - \lfloor
(p/100\%)\,N_{G} \rfloor$ are generated from the complete sets by
randomly removing $P=\lfloor (p/100\%)\,N_{G} \rfloor$ values, which
are used as validation points. For different degrees of thinning
(typically $p=33\%$ and $66\%$), we generate $S=100$ different
sampling configurations. The predictions at the removed points are
calculated and compared with the true values.

\begin{table}[t]
%\vspace{-10pt}
\addtolength{\tabcolsep}{-4.5pt}
\caption{Mean misclassification
rate $\langle F^* \rangle$ $[\%]$ and standard deviation $S_{F^*}$
 for synthetic Gaussian data with anisotropic Mat\'{e}rn covariance obtained by the DGC, KNN
and FKNN algorithms.} \label{tab:synt_class} \vspace{4pt}
\begin{small}
%\begin{center}
\begin{tabular}{|c|ccc|ccc|ccc|ccc|}
\hline
Levels  & \multicolumn{6}{c}{$N_c=8$}  & \multicolumn{6}{|c|}{$N_c=16$} \\
\hline
$p[\%]$ & \multicolumn{3}{c|}{33}  & \multicolumn{3}{c|}{66} &
\multicolumn{3}{c|}{33} & \multicolumn{3}{c|}{66}  \\
\hline
Model  & DGC & KNN & FKNN & DGC & KNN & FKNN & DGC & KNN & FKNN & DGC & KNN & FKNN\\
\hline
$\langle F^{*} \rangle$ &18.9& 29.1 & 27.5 & 26.9& 35.6& 34.8 &26.4&51.4& 51.4& 38.6 & 58.1 &57.9\\
${\rm S}_{F^{*}}$ & 1.6 &1.4&1.4&2.2&1.2 &1.2& 2.1&1.6&1.5&2.5&1.1 &1.0\\
\hline
\end{tabular}
%\end{center}
\end{small}
%\vspace{-10pt}
\end{table}

\begin{figure}[b!]
%\vspace{-10pt}
\begin{center}
    \subfigure[Original data]{\includegraphics[scale=0.23]{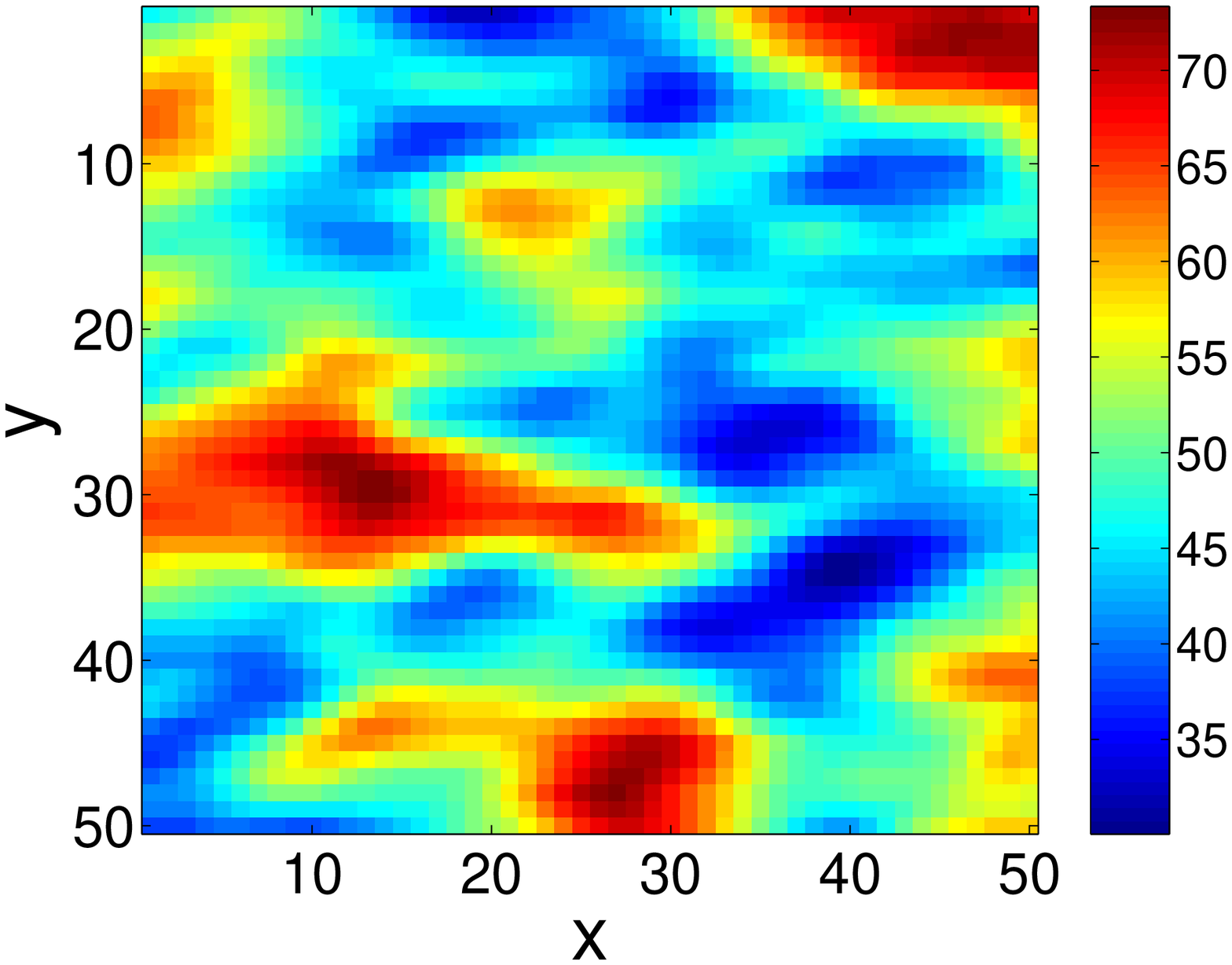}}
    \subfigure[Sample data]{\includegraphics[scale=0.23]{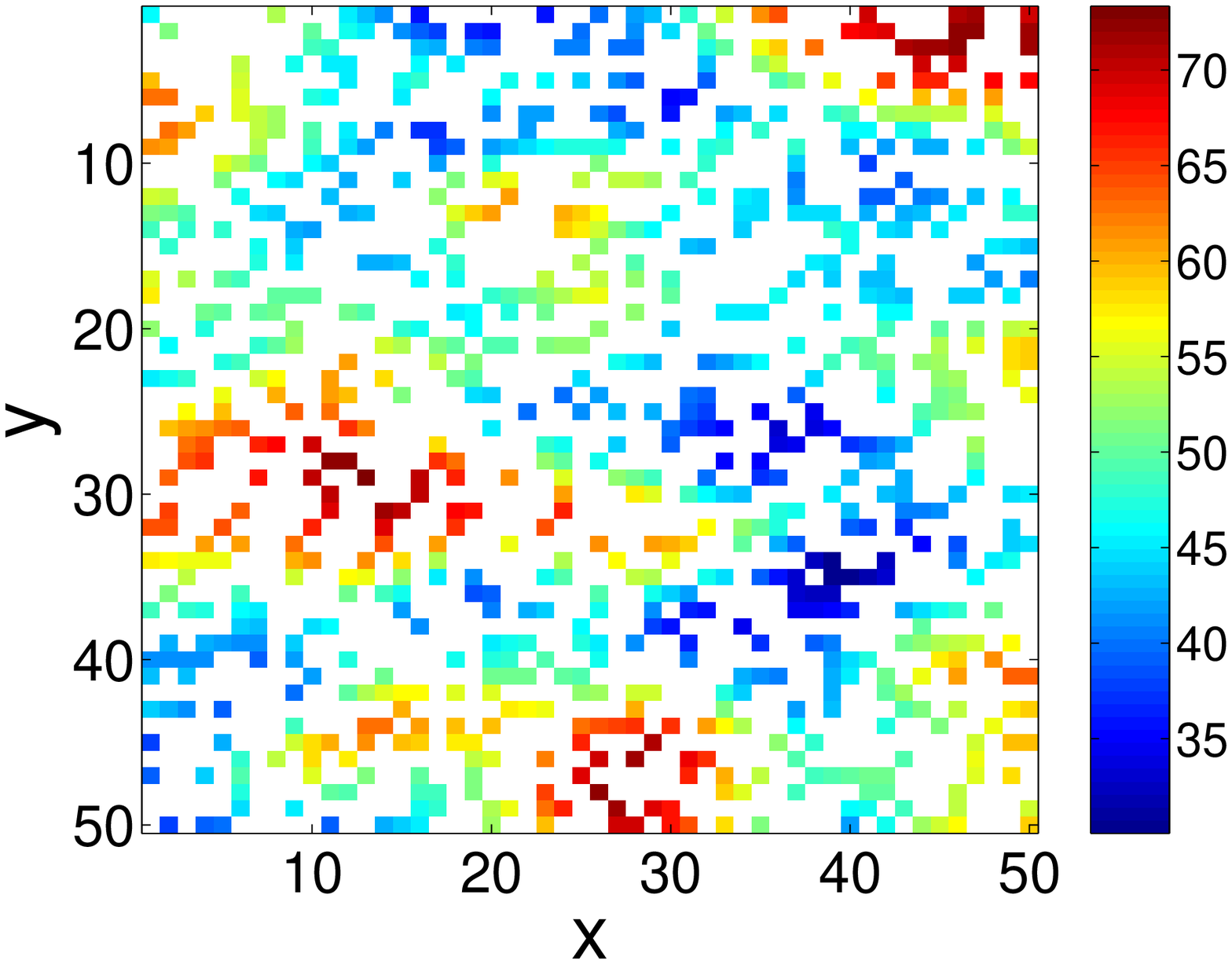}}
    \subfigure[Interpolated data]{\includegraphics[scale=0.23]{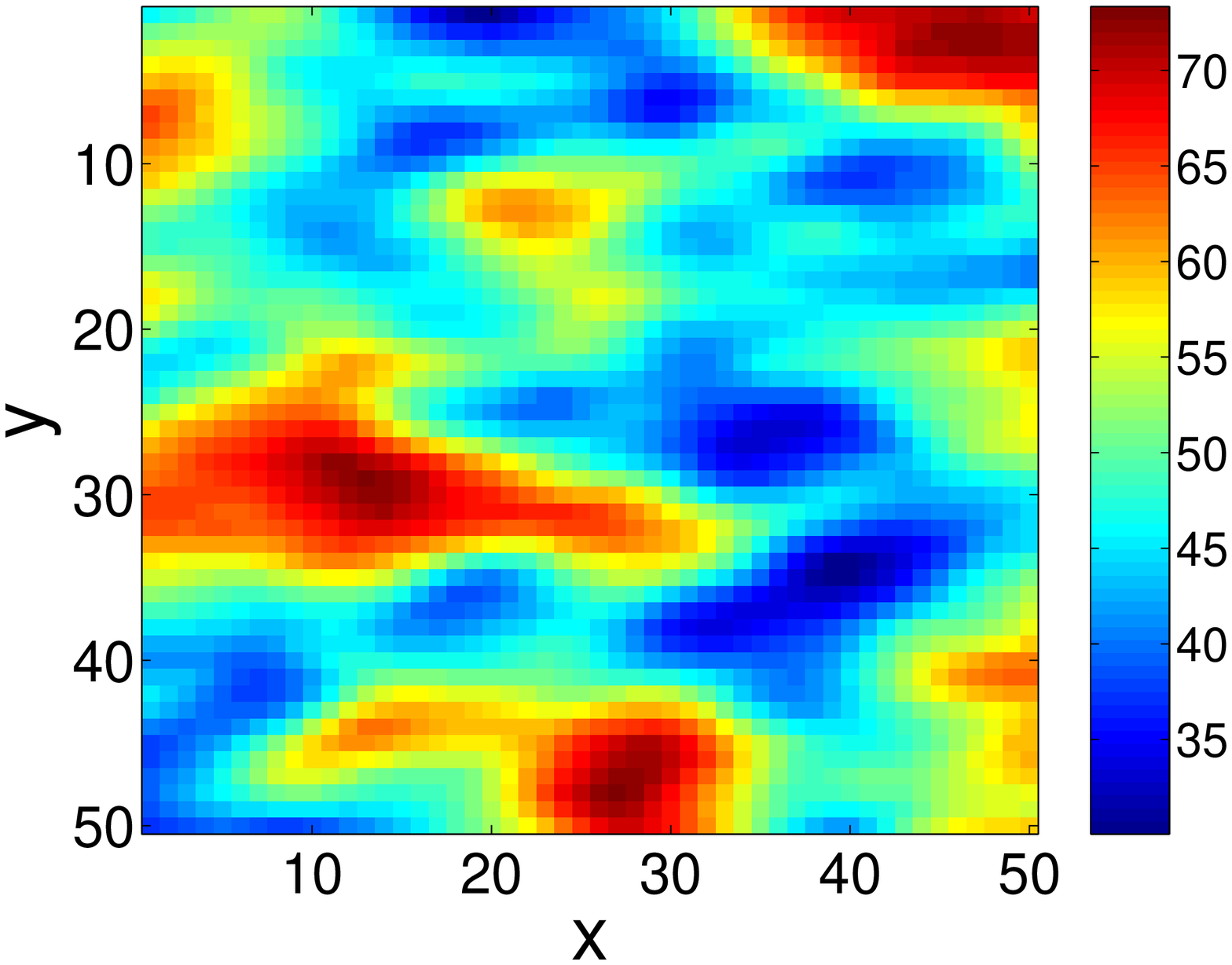}}
    \subfigure[Empirical cdf]{\includegraphics[scale=0.23]{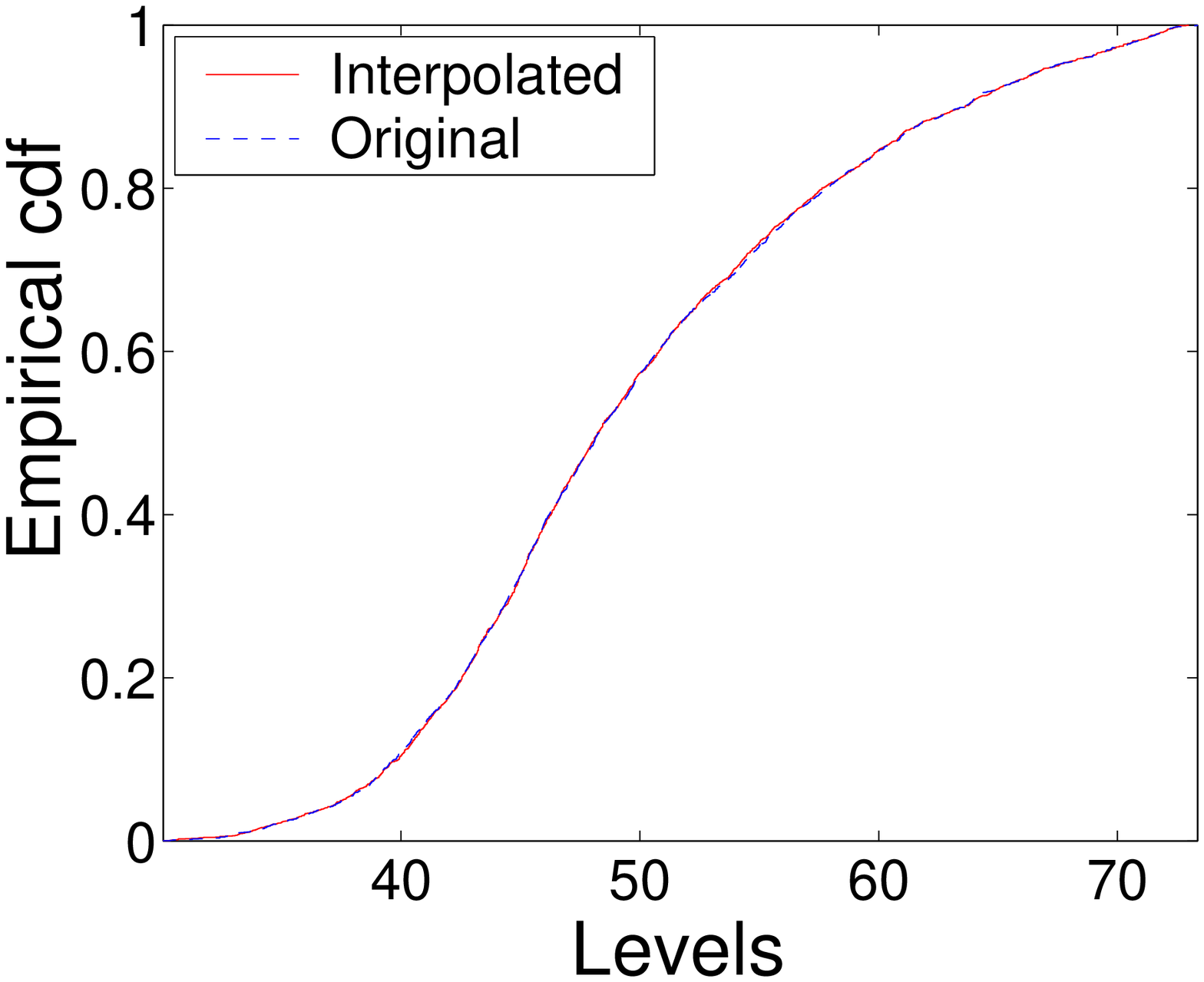}}
    \subfigure[95\% c.i. width]{\includegraphics[scale=0.23]{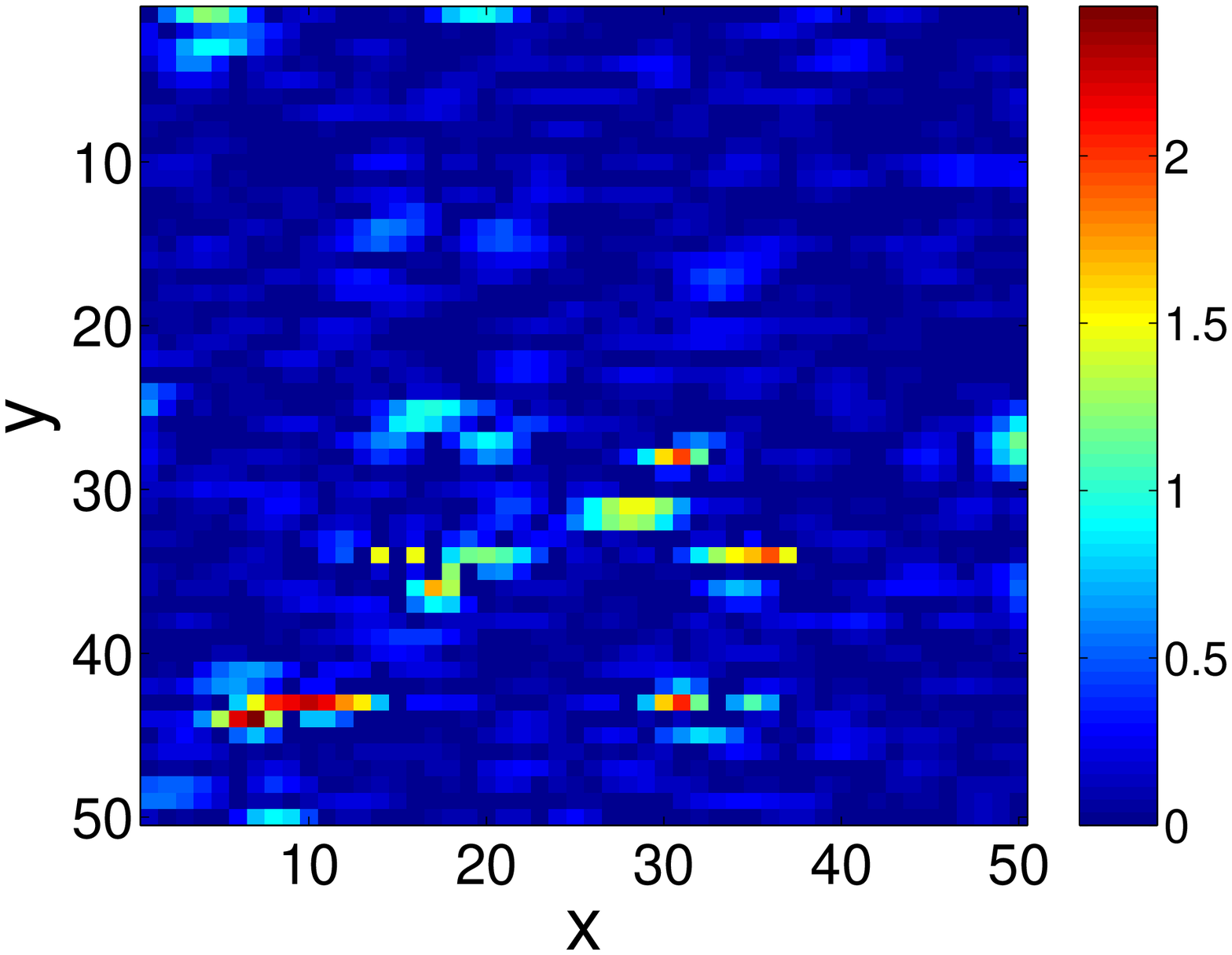}}
    \subfigure[RMSE]{\includegraphics[scale=0.23]{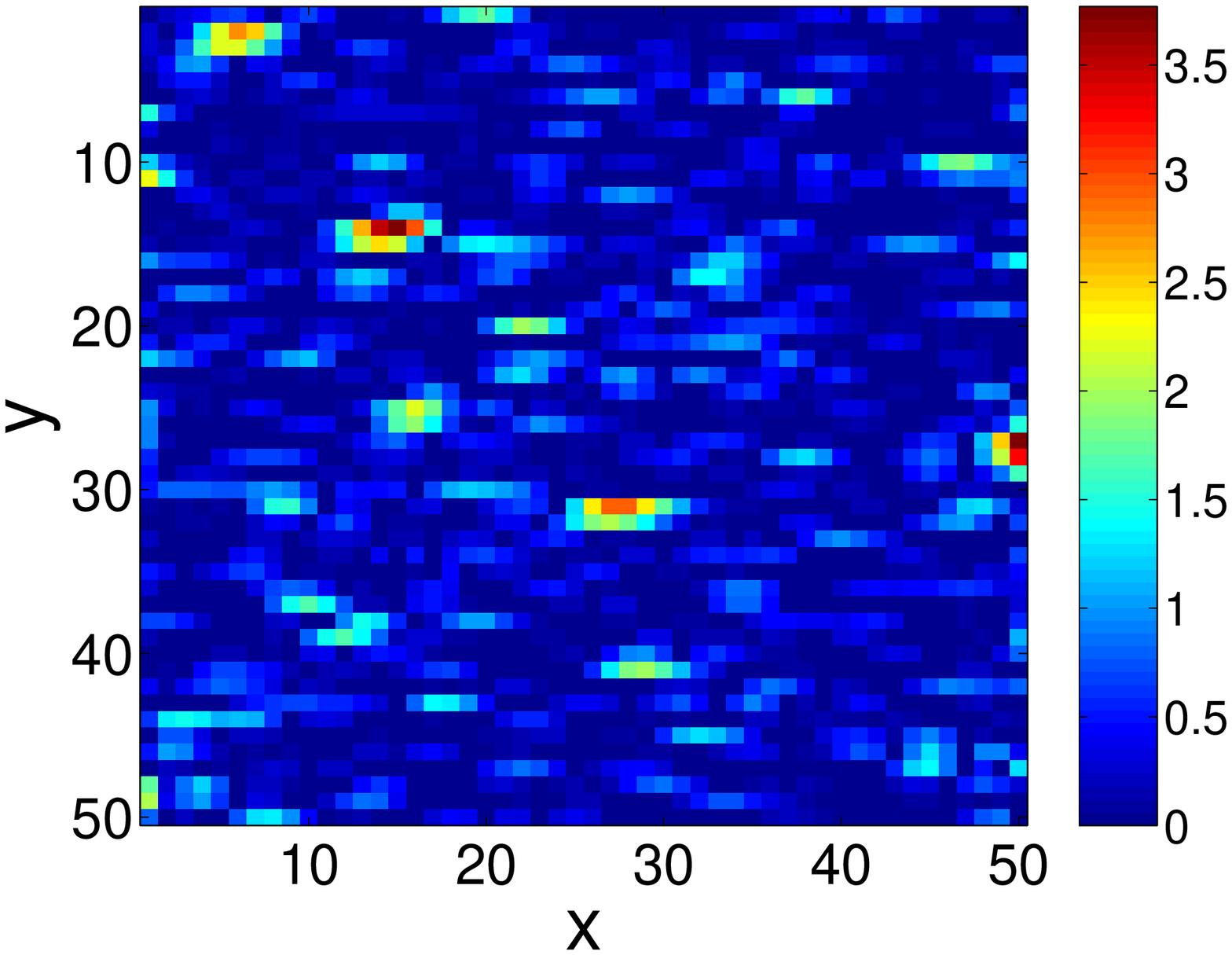}}
    \end{center}
%    \vspace{-10pt}
     \caption{DGC  interpolation results for synthetic Gaussian data with
     anisotropic Mat\'{e}rn covariance based on $M = 100$ simulation runs
     on a single sample generated by $66\%$ thinning. Subfigures include
     (a) original field, (b) thinned sample, (c) interpolated data based
     on the median values from $M$ runs, (d) comparison of the empirical cdfs
     of the original and interpolated data, (e) spatial distribution of
     the $95 \%$ confidence interval (c.i.) widths, and (f) root mean squared
     errors of predictions.}
%     \vspace{-15pt}
    \label{fig:synth}
\end{figure}

The classification results for the synthetic data are summarized in
Table~\ref{tab:synt_class}. The misclassification rate obtained by
DGC is considerably smaller than the KNN and FKNN rates in all
cases, although DGC shows somewhat larger sample-to-sample
fluctuations. The mean CPU time required by DGC ranges between 0.96
and 1.11 seconds and the mean number of Monte Carlo steps between
$10^4$ and $5\times10^4$. The DGC interpolation performance is
evaluated in Table~\ref{tab:synt_int} using $N_c=1000$ classes. In
terms of validation errors (smallest errors and largest $R$), for
the uniformly thinned data ($p=33\%,66\%$) OK ranked best. As
mentioned above, for Gaussian data with known covariance parameters
OK is expected to give optimal predictions. The known directional
correlation lengths also allowed identifying a region of influence
around the prediction points, thus optimizing search neighborhoods
and consequently the OK CPU time. Nevertheless, the OK CPU time was
the highest. For $p=33\%$ the DGC performance ranked second and for
$p=66\%$ it was comparable to BS, with the other models performing
worse than DGC. We note that DGC values in
Tables~\ref{tab:synt_class} and~\ref{tab:synt_int} are based on
$M=1$ simulation run for each of $S=100$ sample realizations.
Increased values of $M$ (e.g. $M=100$) only marginally improved the
validation results.

\begin{figure}[t!]
%\vspace{-10pt}
\begin{center}
    \subfigure[OK interpolation]{\includegraphics[scale=0.23]{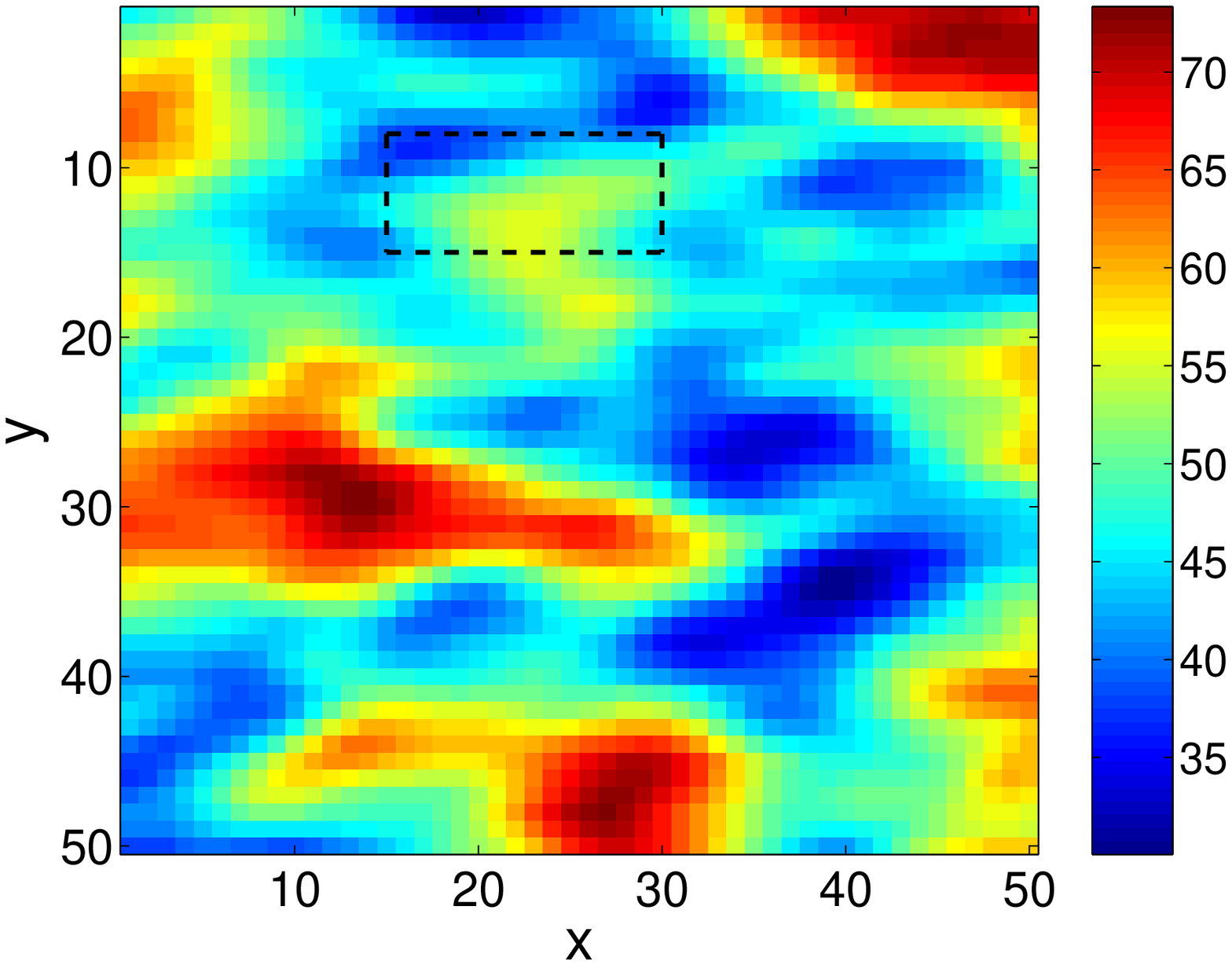}}
    \subfigure[OK variance]{\includegraphics[scale=0.23]{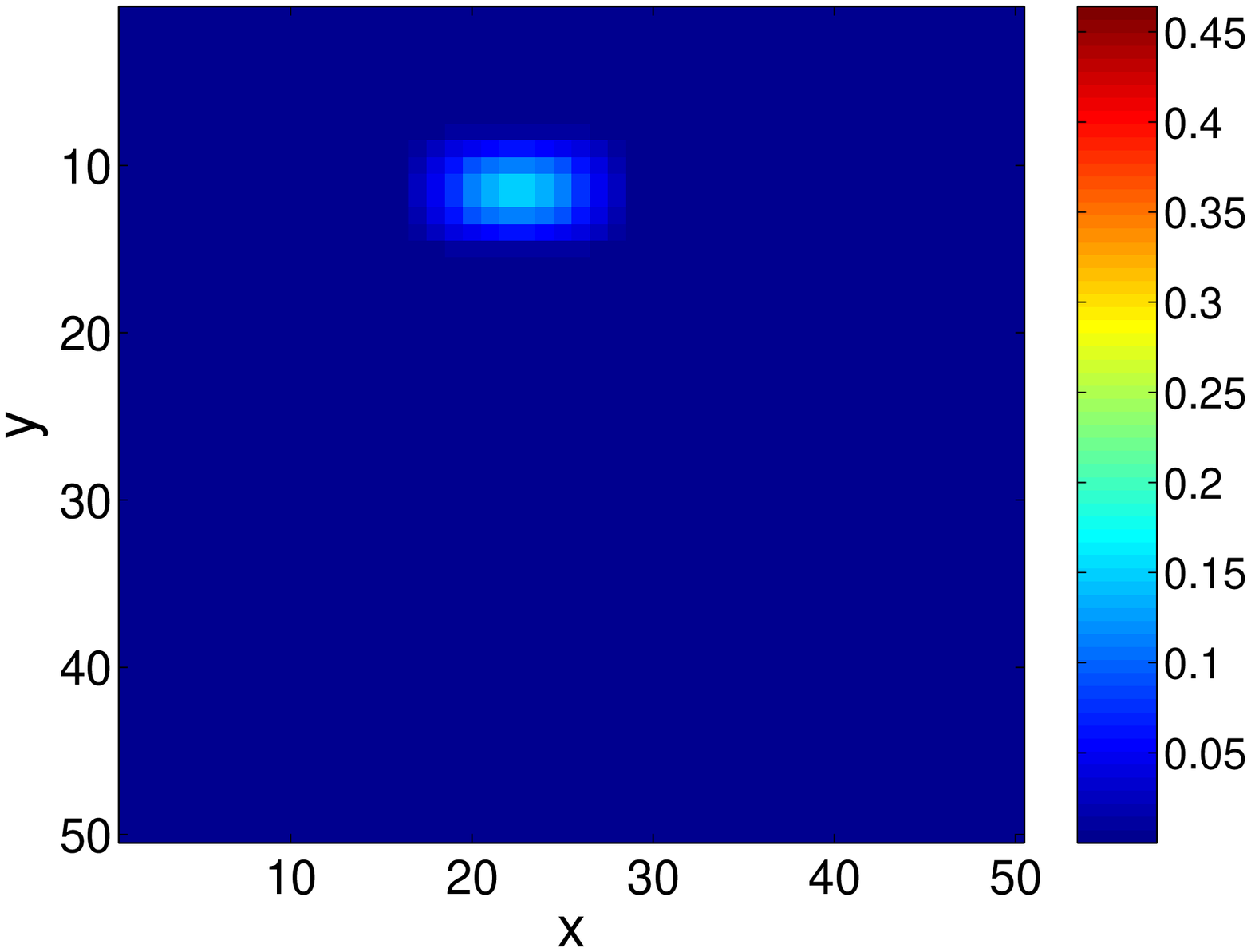}}
    \subfigure[OK absolute error]{\includegraphics[scale=0.23]{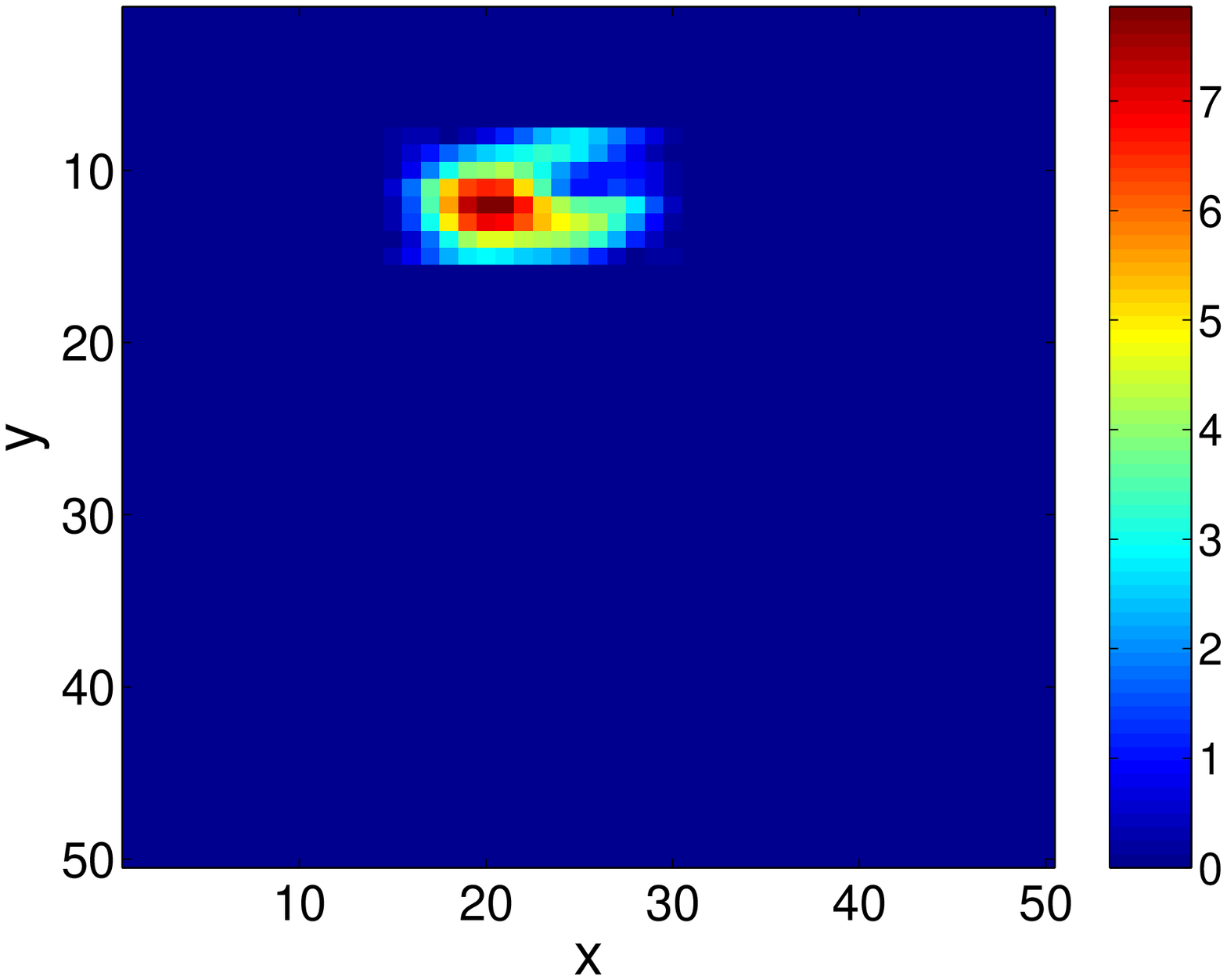}}
    \subfigure[DGC interpolation]{\includegraphics[scale=0.23]{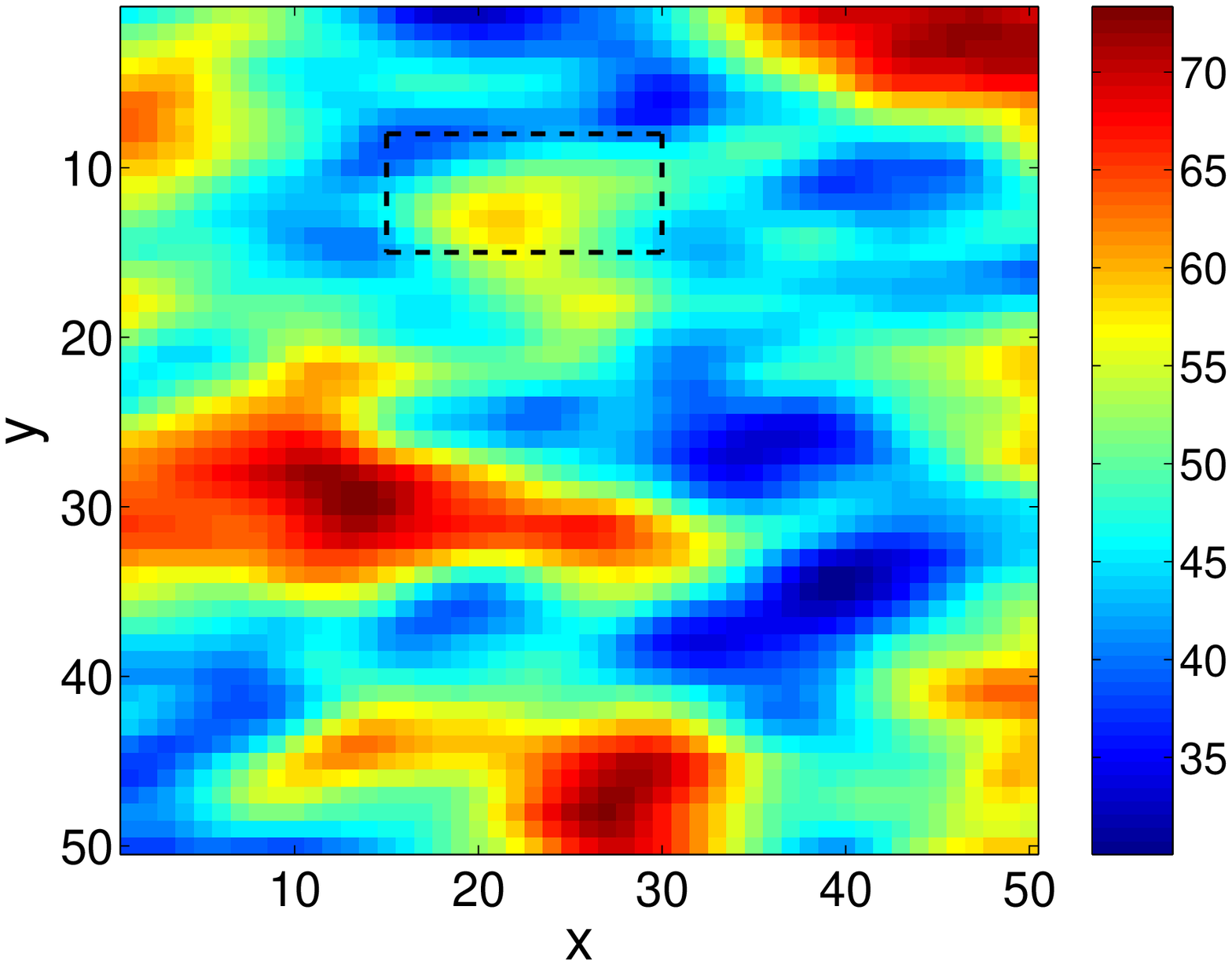}}
    \subfigure[DGC variance]{\includegraphics[scale=0.23]{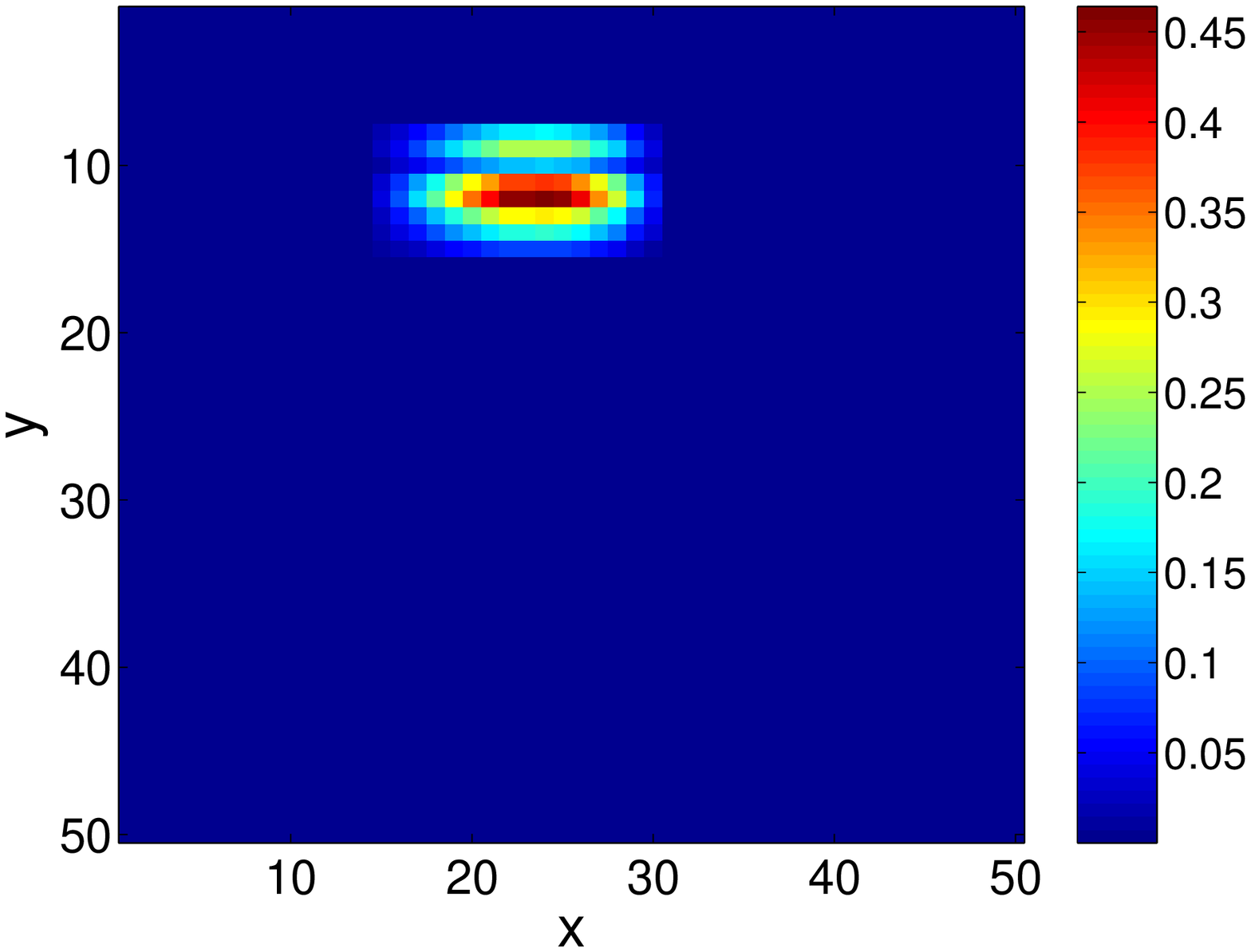}}
    \subfigure[DGC absolute error]{\includegraphics[scale=0.23]{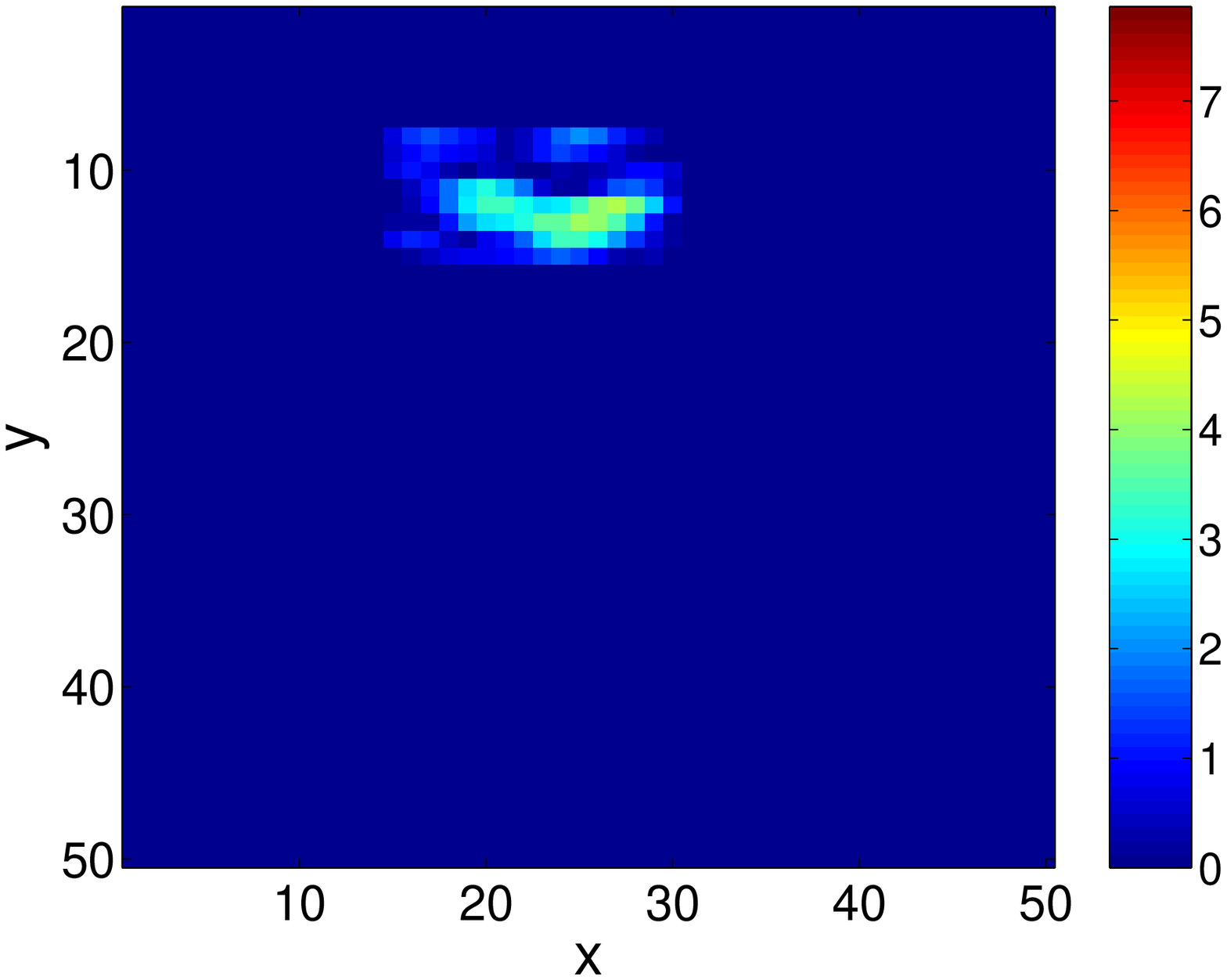}}
    \end{center}
%    \vspace{-10pt}
     \caption{Interpolation results for synthetic Gaussian data with
     anisotropic Mat\'{e}rn covariance. Sample data are generated by removal of a block of
     data in the area marked by the dashed rectangle. The DGC results are based on $M = 100$ simulation
     runs. Subfigures show data interpolated by (a) OK and (d) DGC, variance by (b) OK and (e) DGC, and
     absolute errors by (c) OK and (f) DGC.}
%     \vspace{-15pt}
    \label{fig:synth_blk}
\end{figure}

To account for more realistic patterns of missing data in remote
sensing, e.g. due to cloud cover, we investigate a sample
realization in which a solid block of data (rectangle of $16 \times
8$ pixels) is missing (see Fig.~\ref{fig:synth_blk}). The block is
deliberately chosen to include a small area with extreme values to
test the ability of DGC to predict a ``hotspot''. For this study,
DCG performs better than the other methods. To compensate for using
one sample $(S=1)$, we run $M=100$ simulations. Therefore the DGC
CPU time is considerably higher compared to the (a) and (b) cases
$(M=1)$.

The dramatic increase of the OK CPU time is caused by the augmented
search neighborhood necessary  to span the missing data gap and the
cubic dependence of kriging on the number of points in the search
neighborhood. Generally, the DGC CPU time is proportional to $N_c$
and increases with $p$, reflecting the increased dimension of the
configuration space and number of variables involved in the
optimization. For $p=66\%$ the optimization involves approximately $
10^6$ Monte Carlo steps. As shown in Fig.~\ref{fig:synth}, multiple
simulation runs allow estimating the interpolation uncertainty, with
respect to the subspace of configurations that correspond to local
minima of the objective function~\eqref{cost}.

\begin{table}[t]
%\vspace{-10pt}
\addtolength{\tabcolsep}{-3pt} \caption{Interpolation validation
measures for Gaussian data with anisotropic Mat\'{e}rn covariance,
using (a,b) $S=100$ samples generated by $33\%$ and $66\%$ random
thinning, respectively, and (c) $S=1$ sample generated by removal of
a solid block of data. The DGC uses $N_c=1000$ and the results are
based on $M=1$ simulation run in (a) and (b) and $M=100$ simulation
runs in (c).} \vspace{3pt} \label{tab:synt_int}
\begin{scriptsize}
\resizebox{!}{1.4cm}{
\begin{tabular}{|c|cccccccccccccccccc|}
\hline
 & \multicolumn{3}{c}{MAAE}  & \multicolumn{3}{c}{MARE [\%]} &
 \multicolumn{3}{c}{MAARE [\%]} & \multicolumn{3}{c}{MRASE} &
 \multicolumn{3}{c}{$MR$ [\%]}  & \multicolumn{3}{c|}{$ \langle T_{cpu} \rangle $}   \\
 & (a) & (b) & (c) & (a) & (b) & (c) & (a) & (b) & (c) & (a) & (b) & (c) & (a) & (b) & (c) & (a) & (b) & (c) \\
\hline
DGC &0.17&0.50&1.28&$-$0.01&$-$0.07&1.70&0.35&1.04&2.45&0.33&0.83&1.71&99.93&99.56&98.84&3.16&8.78&166 \\
NN &2.08&2.09&5.01&$-$0.19&$-$0.17&8.12&4.25&4.26&9.77&2.65&2.73&6.24&95.61&95.35&75.52&0.04&0.02&0.08 \\
BL &0.63&1.01&4.95&$-$0.12&$-$0.21&6.49&1.29&2.10&9.41&0.91&1.51&6.07&95.59&95.30&75.52&0.04&0.02&0.06 \\
BC &0.43&0.78&4.92&$-$0.06&$-$0.12&6.82&0.89&1.61&9.32&0.65&1.21&6.09&99.74&99.09&79.00&0.04&0.02&0.06 \\
BS &0.32&0.55&5.09&$-$0.02&$-$0.06&9.19&0.65&1.13&9.61&0.41&0.78&6.30&99.90&99.62&83.45&2.06&0.57&0.49 \\
ID &1.04&1.44&5.14&$-$0.29&$-$0.33&7.33&2.15&2.96&9.85&1.34&1.91&6.22&99.09&97.84&77.06&0.16&0.17&0.06 \\
OK &0.06&0.31&2.49&$-$1E-5&$-$0.01&4.66&0.12&0.65&4.71&0.12&0.49&3.19&99.99&99.85&96.75&31.1&9.15&2546 \\
\hline
\end{tabular}
}
\end{scriptsize}
%\vspace{-10pt}
\end{table}

\subsection{Real Data}
\label{real_data}

\subsubsection{Radioactive Potassium Concentration}
\label{real_data_radio} 
\begin{figure}[t]
%\vspace{-5pt}
\begin{center}
    \subfigure[Original data]{\label{fig:real_radio_a}\includegraphics[scale=0.23]{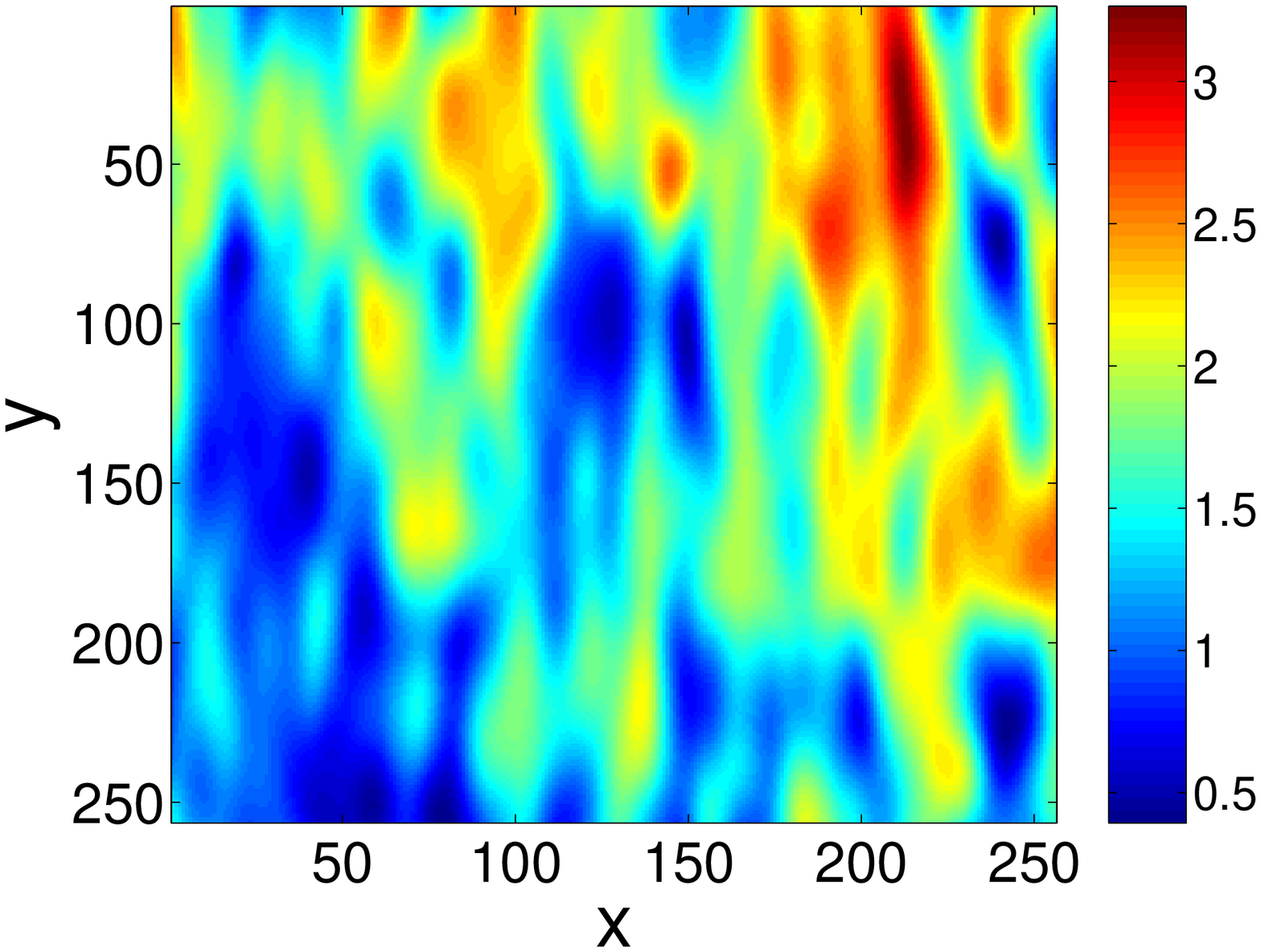}}
    \subfigure[Sample data]{\label{fig:real_radio_b}\includegraphics[scale=0.23]{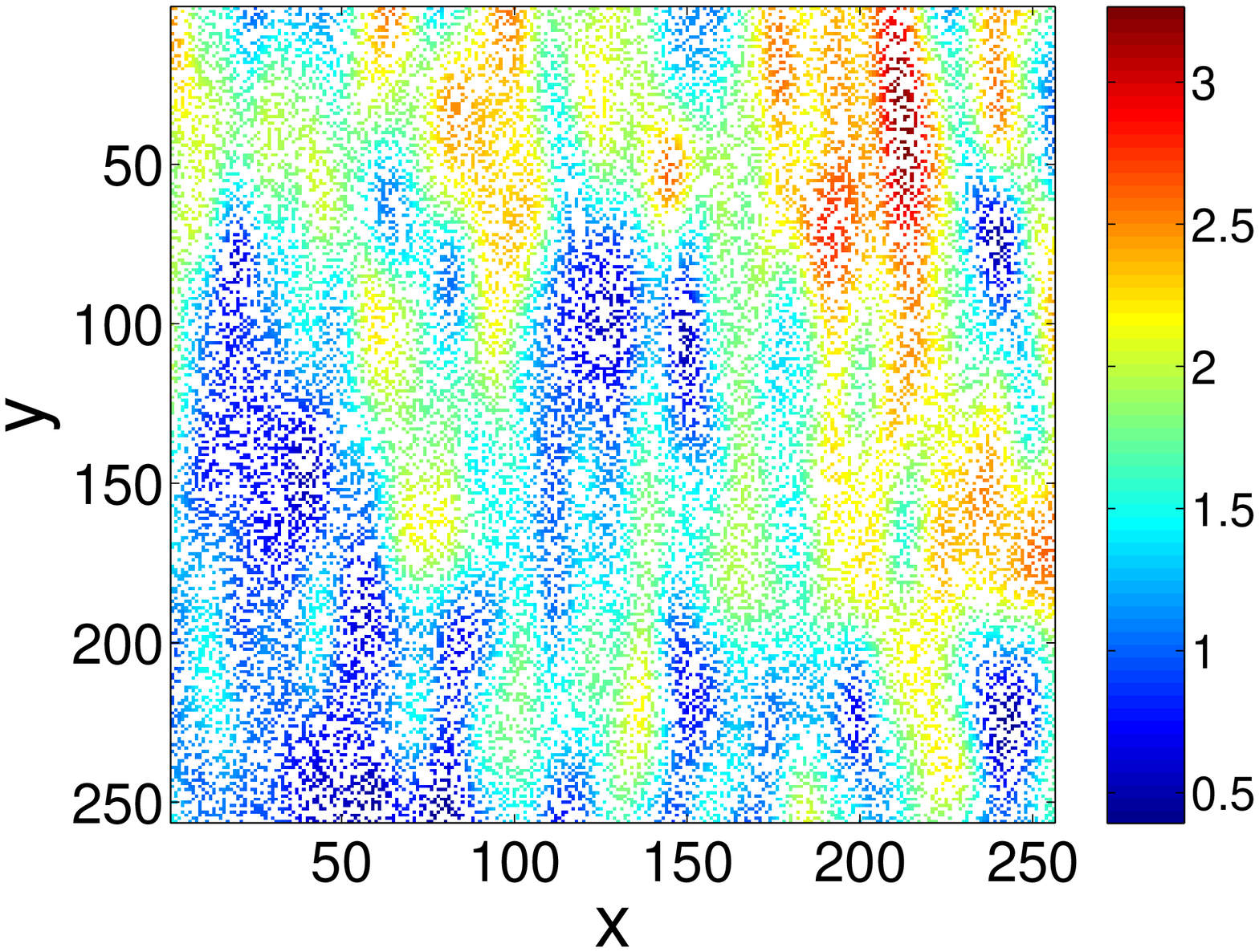}}
    \subfigure[Interpolated data]{\label{fig:real_radio_c}\includegraphics[scale=0.23]{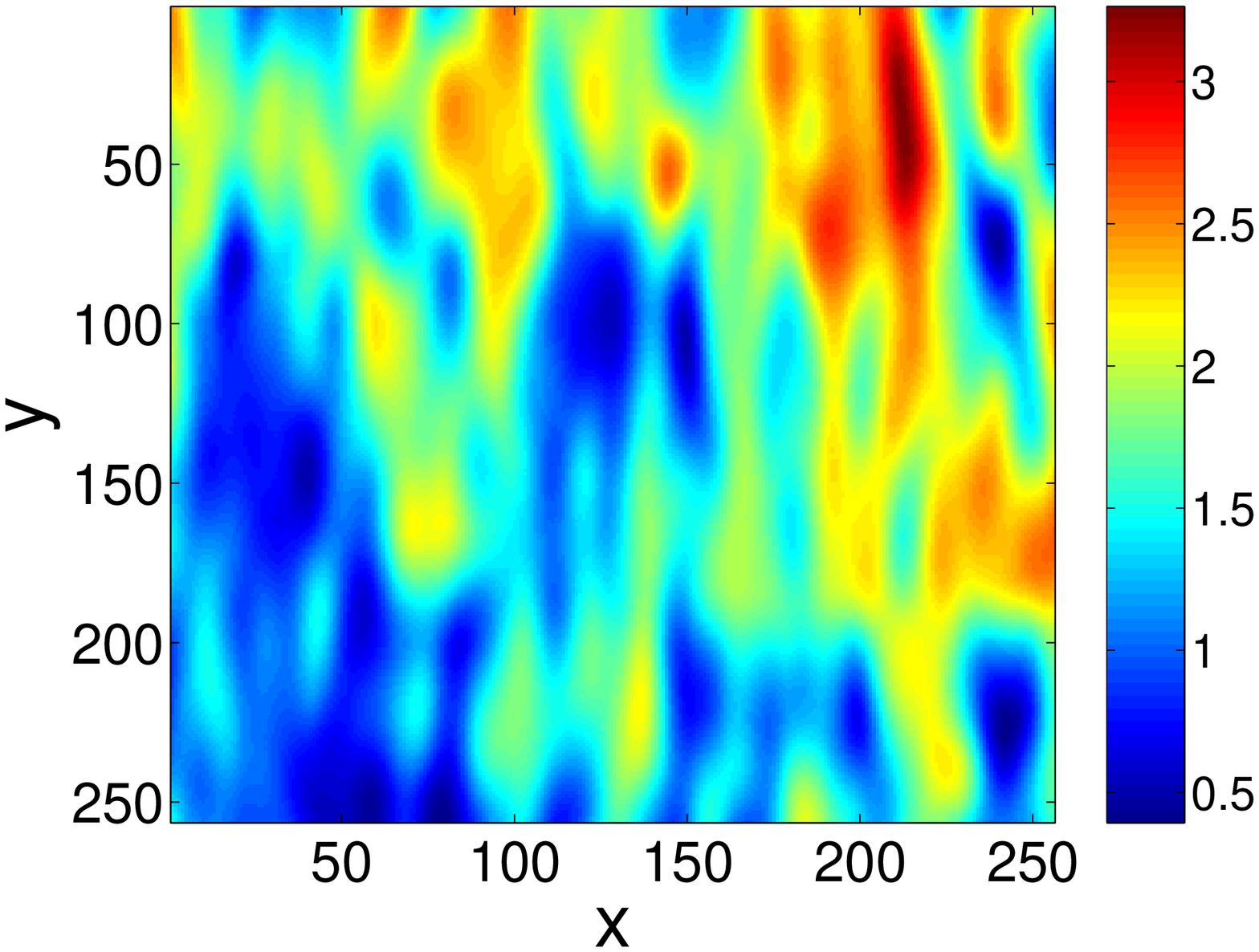}}
    \subfigure[Empirical cdf]{\label{fig:real_radio_d}\includegraphics[scale=0.23]{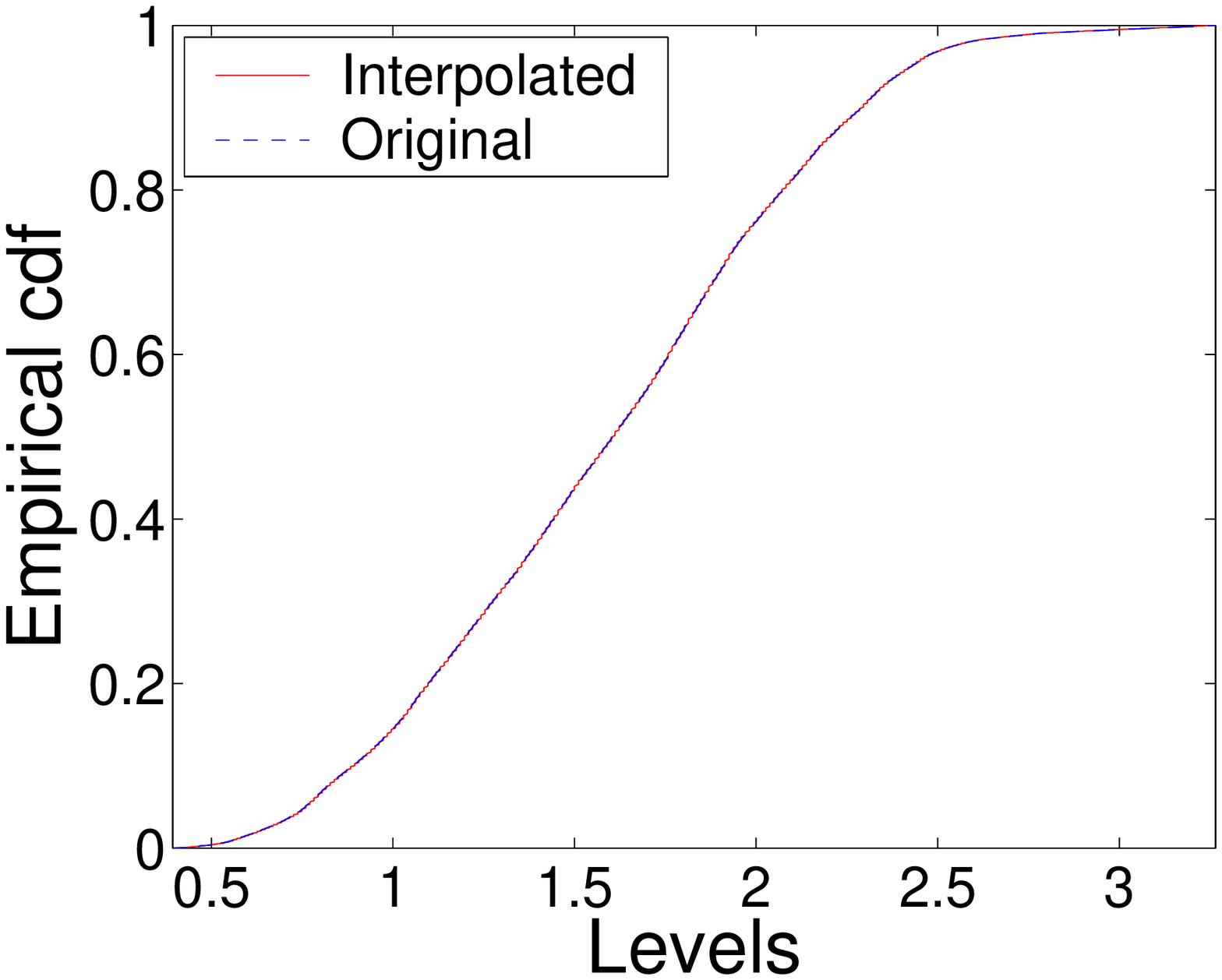}}
    \subfigure[95\% c.i. width]{\label{fig:real_radio_e}\includegraphics[scale=0.23]{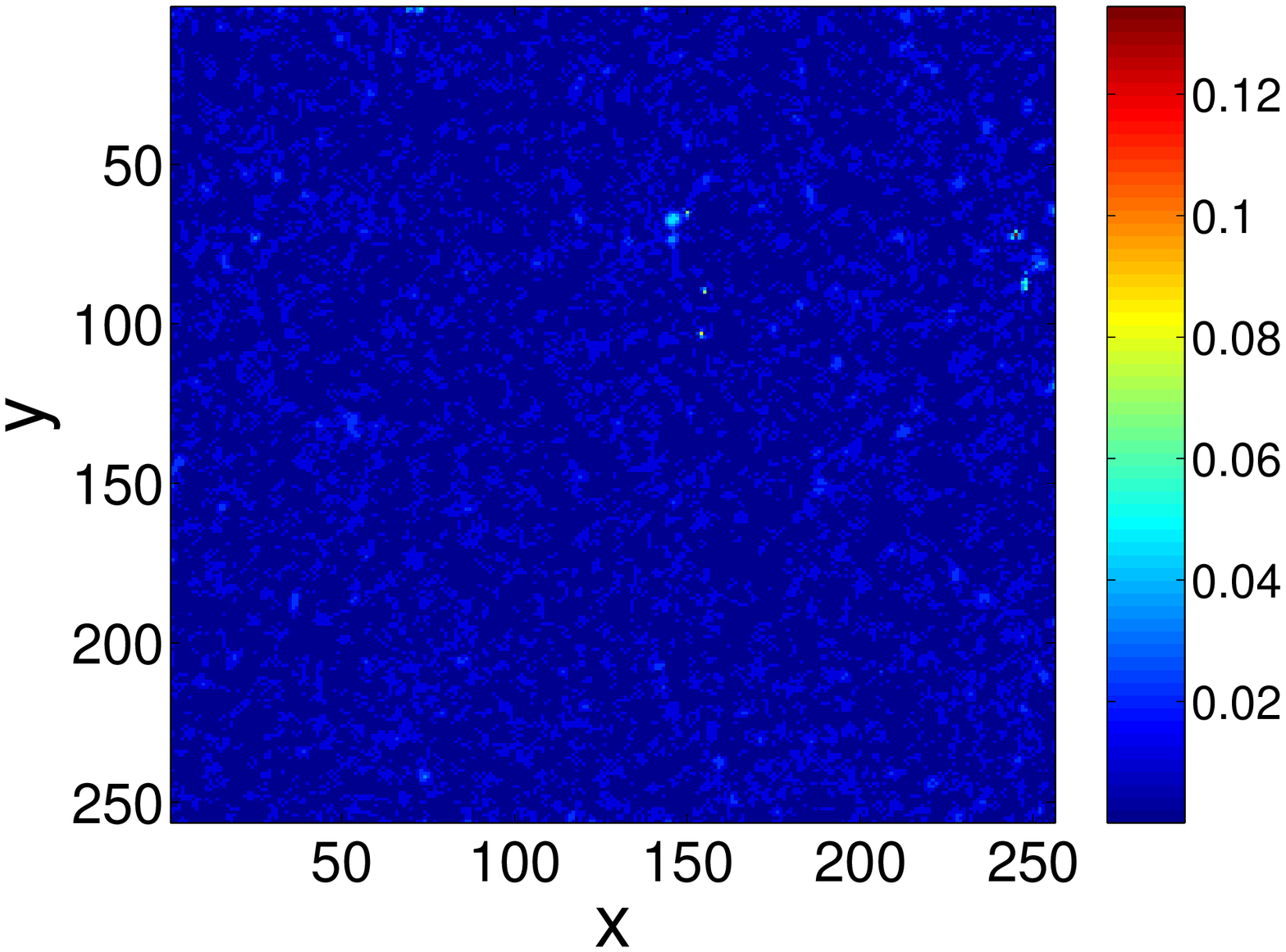}}
    \subfigure[RMSE]{\label{fig:real_radio_f}\includegraphics[scale=0.23]{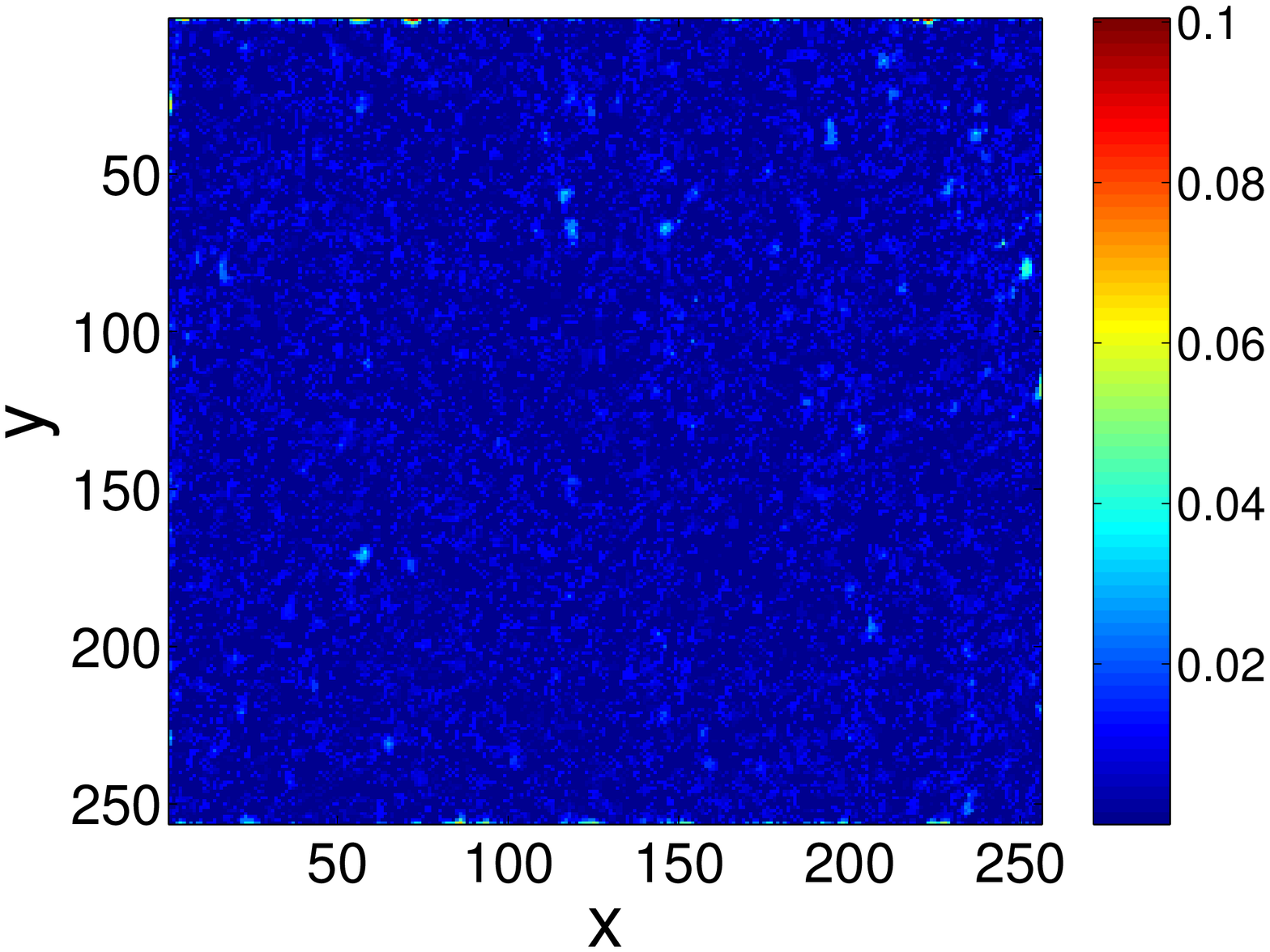}}
    \end{center}
%    \vspace{-10pt}
\caption{DGC interpolation of the radioactivity data obtained  from
$M=100$ simulation runs using one sample set generated by $66\%$
thinning. The plots are as those in Fig.~\ref{fig:synth}.}
%\vspace{-10pt}
    \label{fig:real_radio}
\end{figure}

The first real data set represents soil
concentration of radioactive potassium measured by gamma-ray spectrometry 
over part of Canada~\citep*{Anonymous08}, on a grid with $N_{G}=256 \times 256$
nodes extending in latitude from 56S to 57N and in longitude from
$-$100W to $-$98E, with a resolution of 250 m. The data have been
preprocessed to correct for background and airplane flight height.
The potassium concentrations are in units of \% and their summary
statistics are as follows: $N_{G}=65536$, $z_{\min}=0.39$,
$z_{\max}= 3.26$, $\bar{z}=1.60$, $z_{0.50}= 1.61$,
$\sigma_{z}=0.52$, skewness coefficient equal to $0.10$, and
kurtosis coefficient equal to $2.45$. A plot of the data in Fig.
\ref{fig:real_radio_a} displays clear signs of anisotropy. Samples
${\bf Z}(G_{s})$ were generated from the original data by random
thinning with $p=33\%$ and $66\%$.

\begin{figure}[h!]
\begin{center}
    \subfigure[Objective function evolution]{\label{fig:pota_cost}
    \includegraphics[width=0.45\textwidth]{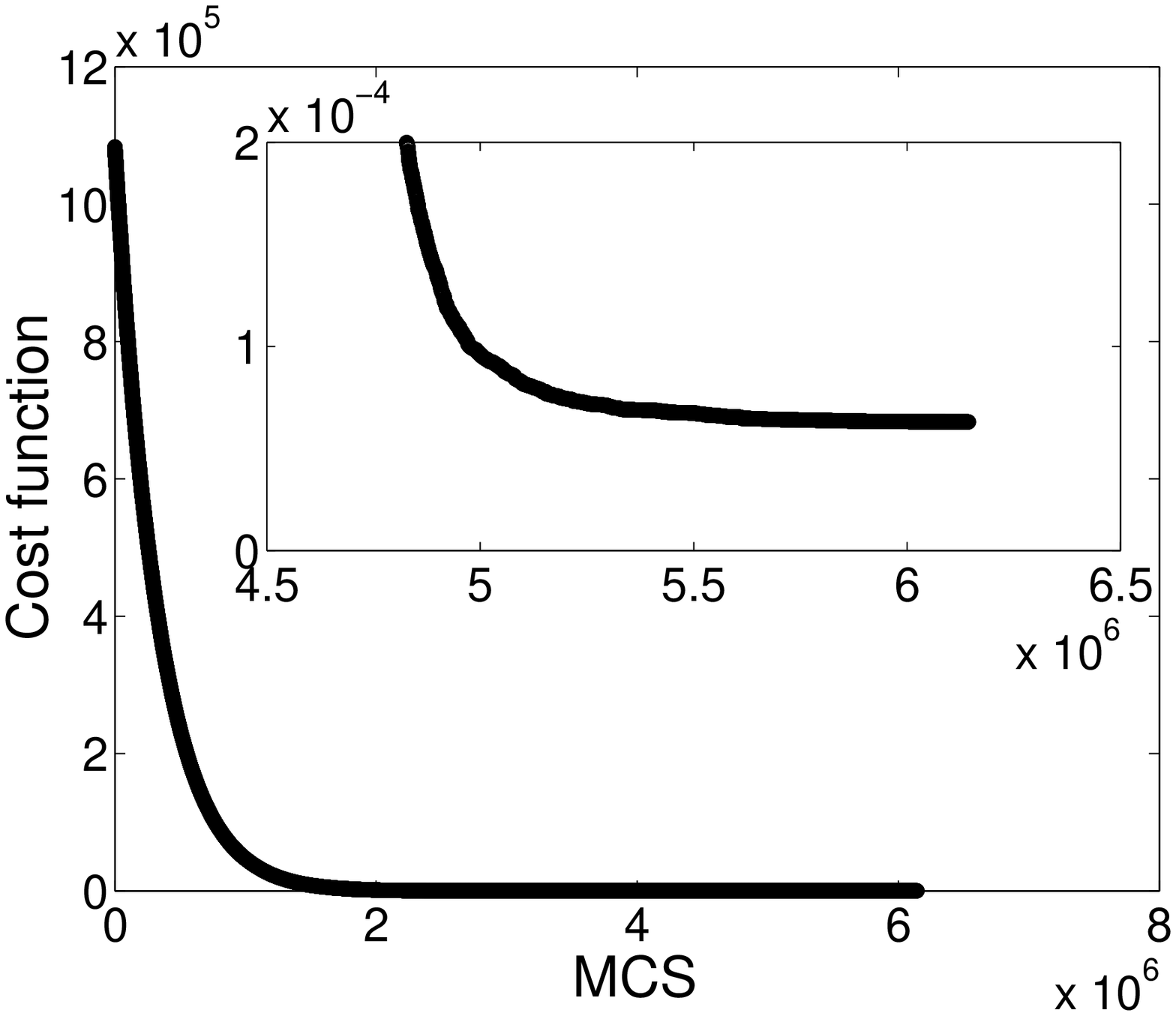}}
    \subfigure[Residual objective function]{\label{fig:pota_cf_hist}
    \includegraphics[width=0.47\textwidth]{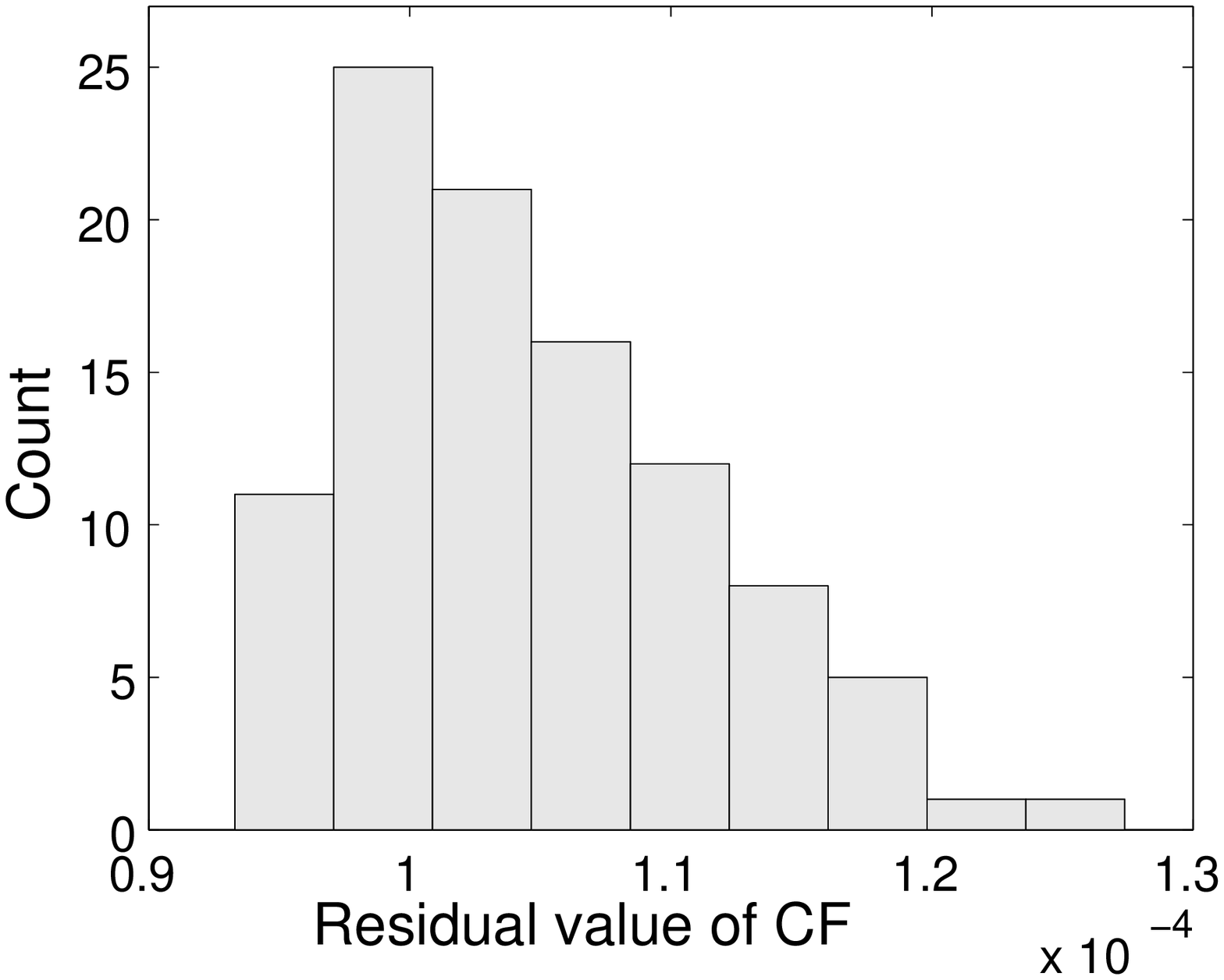}}
\end{center}
   \caption{Left: Evolution of the objective (cost) function~\eqref{cost}
   for potassium concentration versus the number of
   Monte Carlo steps.
   Inset focuses on the convergence to the optimum.
   Right: Histogram of the objective function residual values obtained from
   $100$ different runs. }
    \label{fig:pota_cost_tot}
\end{figure}

Classification $(N_{c}=8, 16)$ and interpolation $(N_{c}=1500)$
results for $p=33\%$ are listed in Table \ref{tab:real_pred}. The
prediction performance of DGC is superior to other models except for
the BS. The outstanding performance of the latter is likely due to
the smooth spatial variation of the radioactivity data. An example
of prediction results based on $M=100$ simulation runs for one
sample realization with $p=66\%$ is shown in
Fig.~\ref{fig:real_radio}. The DGC classification  CPU time with
$N_c=8,16$ was 2.6 and 3.3 seconds respectively. The  computational
time for DGC interpolation is comparable to that of BS, but one
order higher than NN, BL, and BC times. The limiting factor in DGC
are the MC simulations that involve up to $\sim 10^7$ Monte Carlo
steps (see Fig.~\ref{fig:pota_cost}). The histogram in
Fig.~\ref{fig:pota_cf_hist} gives the distribution of  the objective
function residuals for 100 accepted configurations and verifies that
all of them correspond to small $(<2\times 10^{-4})$ values.

\begin{table}[h!]
%\vspace{-5pt}
\addtolength{\tabcolsep}{-4.5pt} \caption{Classification
($N_c=8,16$) and interpolation $(N_c=1500)$ results  for the
radioactivity data with $p=33 \%$ thinning.  DGC is compared with
k-nearest neighbor (KNN) and fuzzy k-nearest neighbor (FKNN) models
for classifications, and with the nearest neighbor (NN), bilinear
(BL), bicubic (BC), biharmonic spline (BS) and inverse distance (ID)
models for interpolation.} \label{tab:real_pred}
\begin{small}
%\begin{center}
\begin{tabular}{|c|cc|cc|cccccc|}
\hline & \multicolumn{4}{c|}{{\bf Classification}} &
\multicolumn{6}{c}{{\bf Interpolation}} \\
\hline
 &  \multicolumn{2}{c}{$N_c=8$}  &
\multicolumn{2}{|c|}{$N_c=16$} &
\multicolumn{6}{|c|}{$N_c=1500$} \\
Model & $\langle F^{*} \rangle$ & ${\rm S}_{F^{*}}$ & $\langle F^{*} \rangle$
& ${\rm S}_{F^{*}}$ & MAAE  & MARE [\%] & MAARE [\%] & MRASE & $MR$ [\%]
& $ \langle T_{cpu} \rangle $\\
\hline
DGC  &4.01&0.21&8.64&0.28&9.4e$-$4 &$-$2.2e$-$3 &6.8e$-$2 & 1.56e$-$3& 100.00& 39.94\\
KNN  &4.96&0.16&11.44&0.22&- &- &- & -& -& -\\
FKNN  &4.22&0.16&10.14&0.23&- &- & &- & -& -\\
\hline
NN  &-&-&-&-&2.3e$-$2 &-0.165 &1.64 & 3.5e$-$2& 99.78& 1.41\\
BL  &-&-&-&-&3.7e$-$3 &-5.9e$-$2 &0.27 & 5.9e$-$3& 99.78& 1.40\\
BC  &-&-&-&-&1.7e$-$3 &-2.2e$-$2 &0.12 & 2.9e$-$3 & 100.00& 1.46\\
BS  &-&-&-&-&4.7e$-$4 &-1.2e$-$3 &3.4e$-$2 & 7.6e$-$4& 100.00& 32.82\\
ID  &-&-&-&-&1.3e$-$2 &-0.160 & 0.90 &1.7e$-$2 & 99.95& 183.64\\
\hline
\end{tabular}
%\end{center}
\end{small}
%\vspace{-10pt}
\end{table}

\subsubsection{Ozone Layer Thickness}
\label{real_data_ozone} The second real-world data set represents
daily column ozone measurements on June 1, 2007 \citep*{acker07}.
The data are on a $1^\circ \times 1^\circ$ grid with $N_{G}=180
\times 360$ nodes extending in latitude from 90S to 90N and in
longitude from 180W to 180E. The data set includes naturally missing
(and therefore unknown) values. The data are in Dobson units with
the following summary statistics: $N=48501$, $z_{\min}=158$,
$z_{\max}= 596$, $\bar{z}=311.58$, $z_{0.50}= 308$,
$\sigma_{z}=46.05$, skewness coefficient equal to $0.31$, and
kurtosis coefficient equal to $2.30$. The gaps are mainly due to
limited coverage on the particular day, generating conspicuous
stripes of missing values in the south-north direction. Since the
true values at these locations are not known, validation measures
are not evaluated. Instead, the interpolation quality is assessed
empirically, based on the visual continuity between  the observed
data and the predictions.

\begin{figure}[h!]
%\vspace{-10pt}
\begin{center}
    \subfigure[Gappy ozone data]{\label{fig:ozone_DGC_a}\includegraphics[scale=0.23]{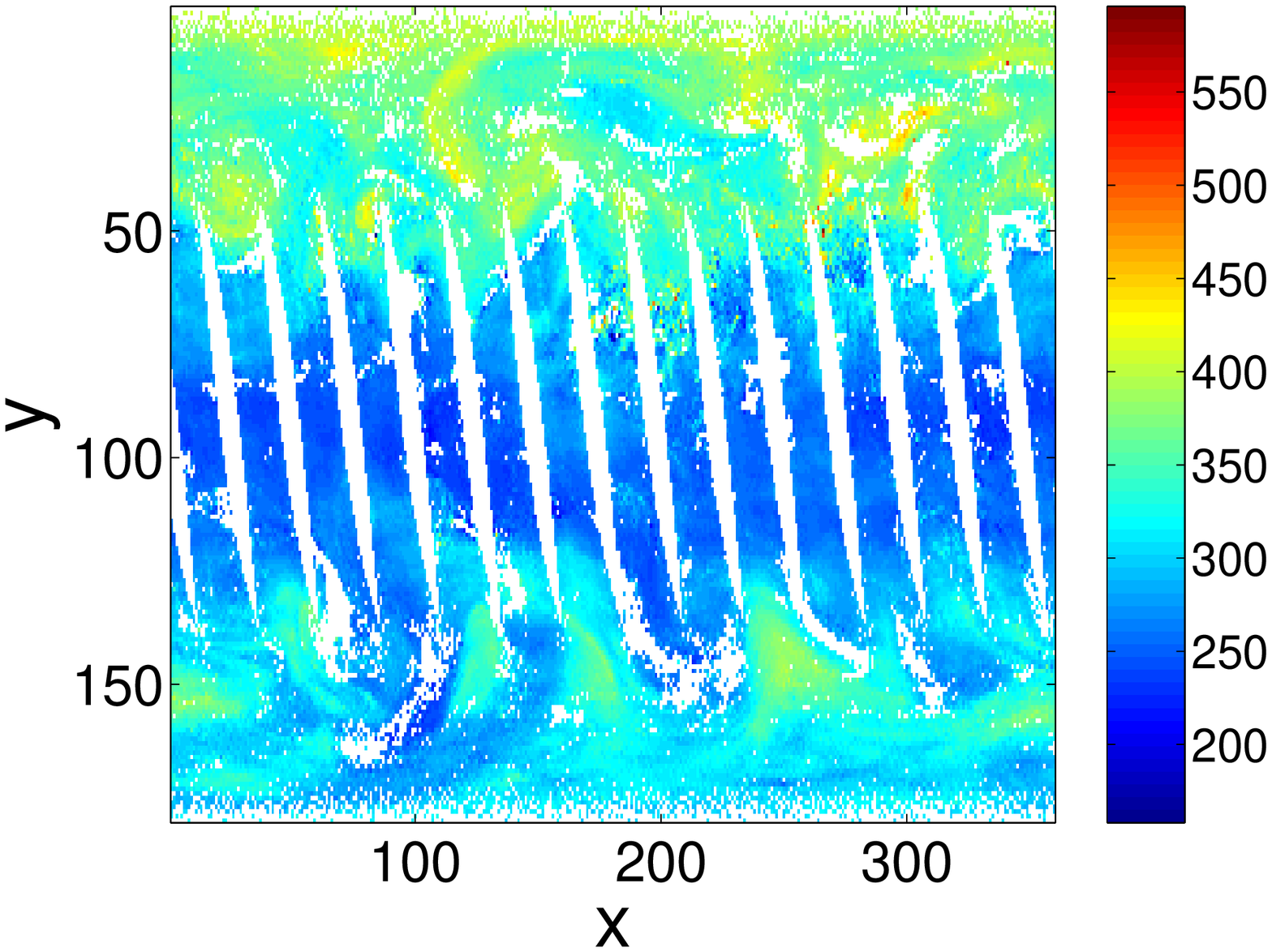}}
    \subfigure[DGC  interpolation]{\label{fig:ozone_DGC_b}\includegraphics[scale=0.23]{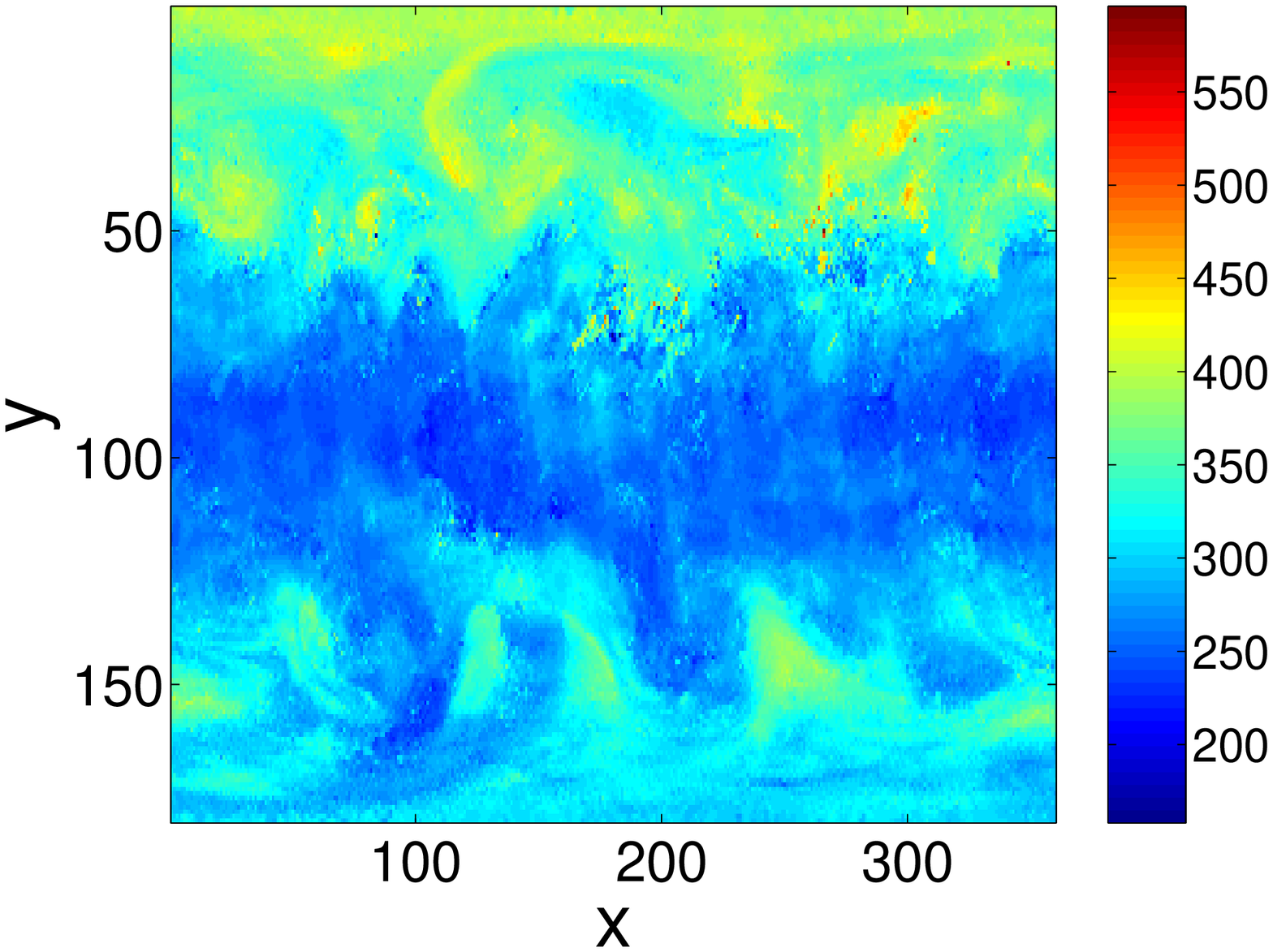}}
    \subfigure[95\% c.i. width]{\label{fig:ozone_DGC_c}\includegraphics[scale=0.23]{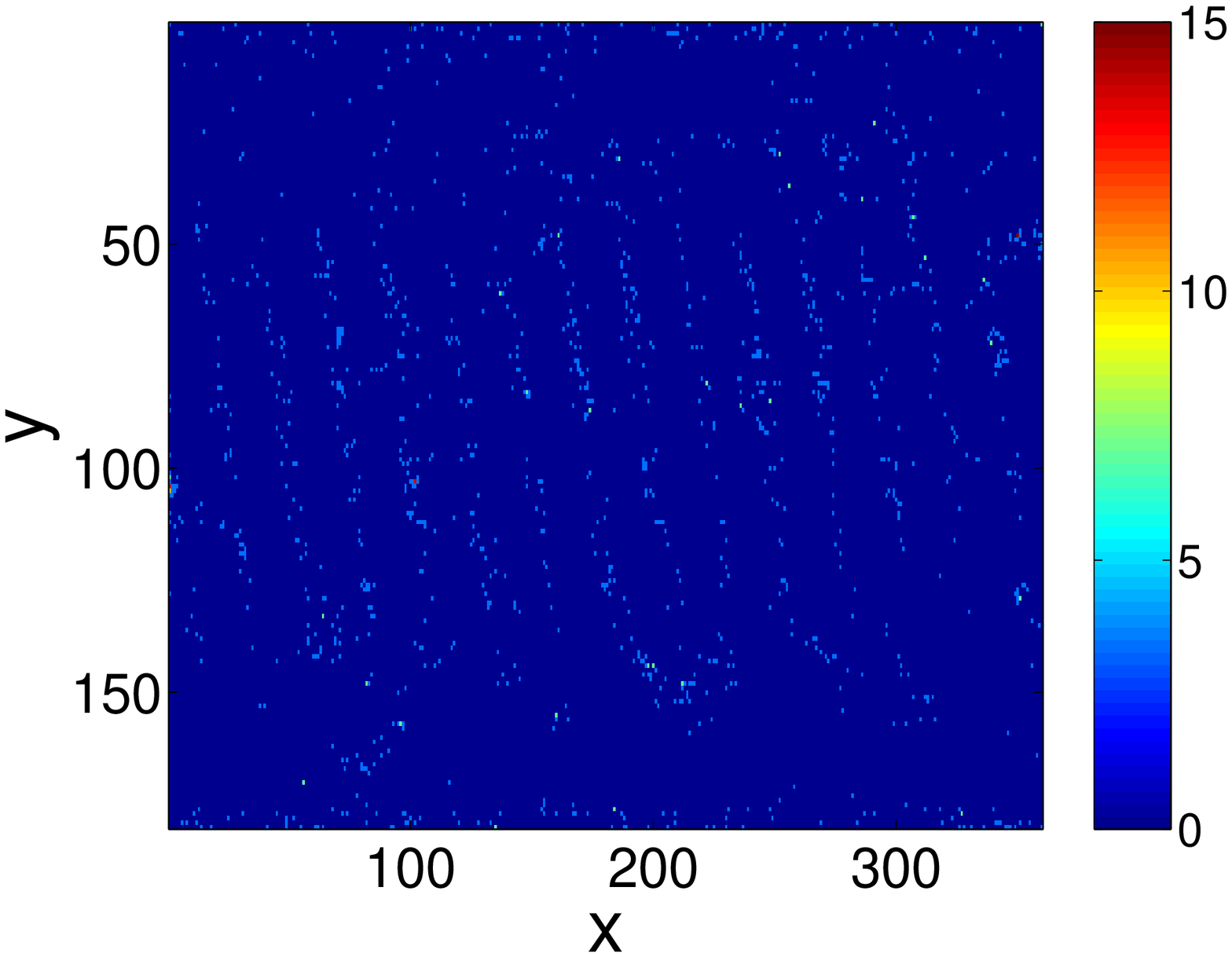}}
    \end{center}
%    \vspace{-10pt}
     \caption{DGC  interpolation results for ozone data,
     obtained from $M=100$ simulation runs on
     one set of the original data with missing values: Original data (a),
      interpolated data based on the median values from $M$ runs (b), and
     spatial distribution of the $95 \%$ confidence interval (c.i.) widths (c).}
%     \vspace{-15pt}
    \label{fig:ozone_DGC}
\end{figure}

\begin{figure}[h!]
%\vspace{-10pt}
\begin{center}
    \subfigure[NN interpolation]{\includegraphics[scale=0.23]{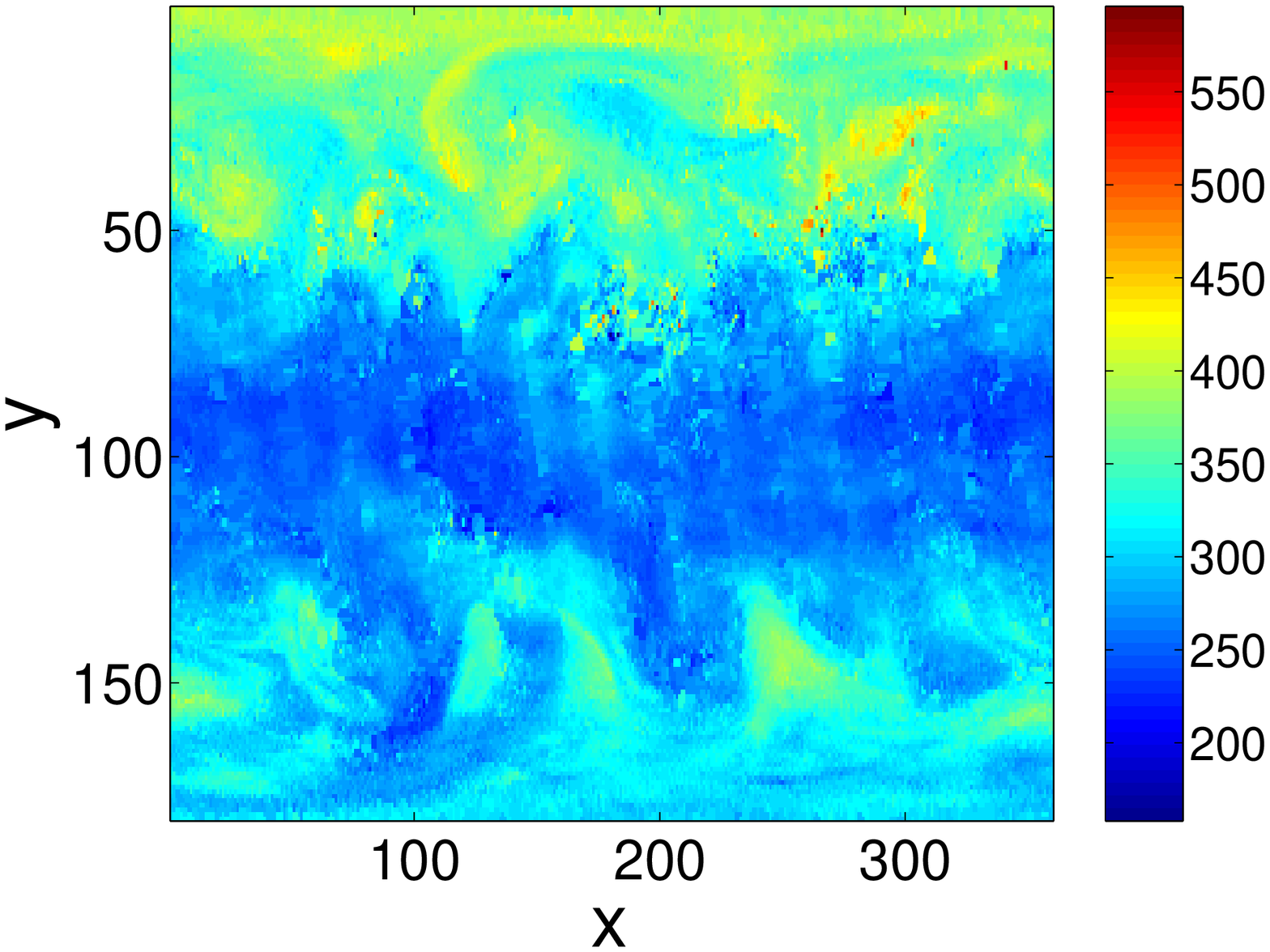}}
    \subfigure[ID interpolation]{\includegraphics[scale=0.23]{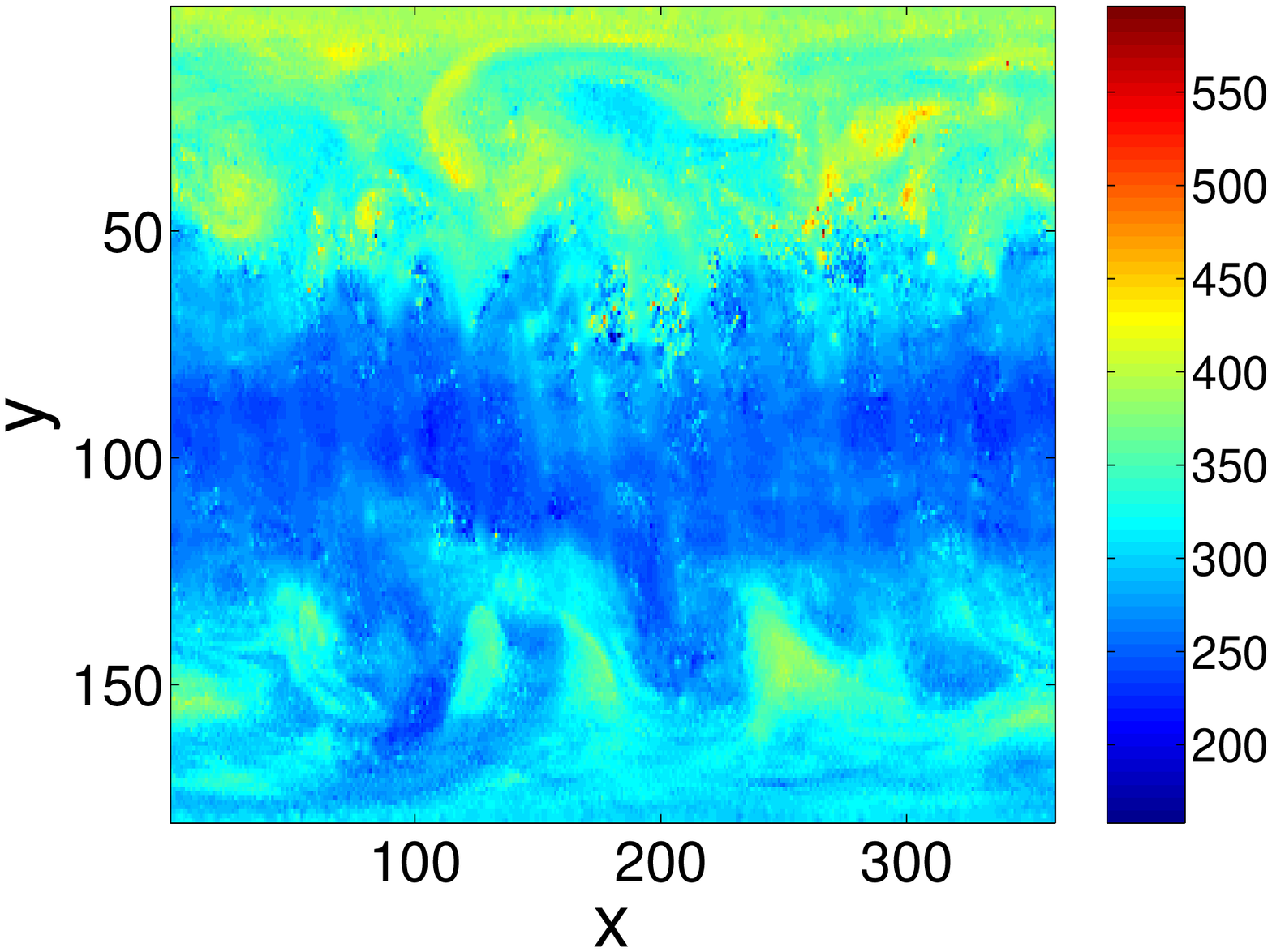}}
    \subfigure[BS interpolation]{\includegraphics[scale=0.23]{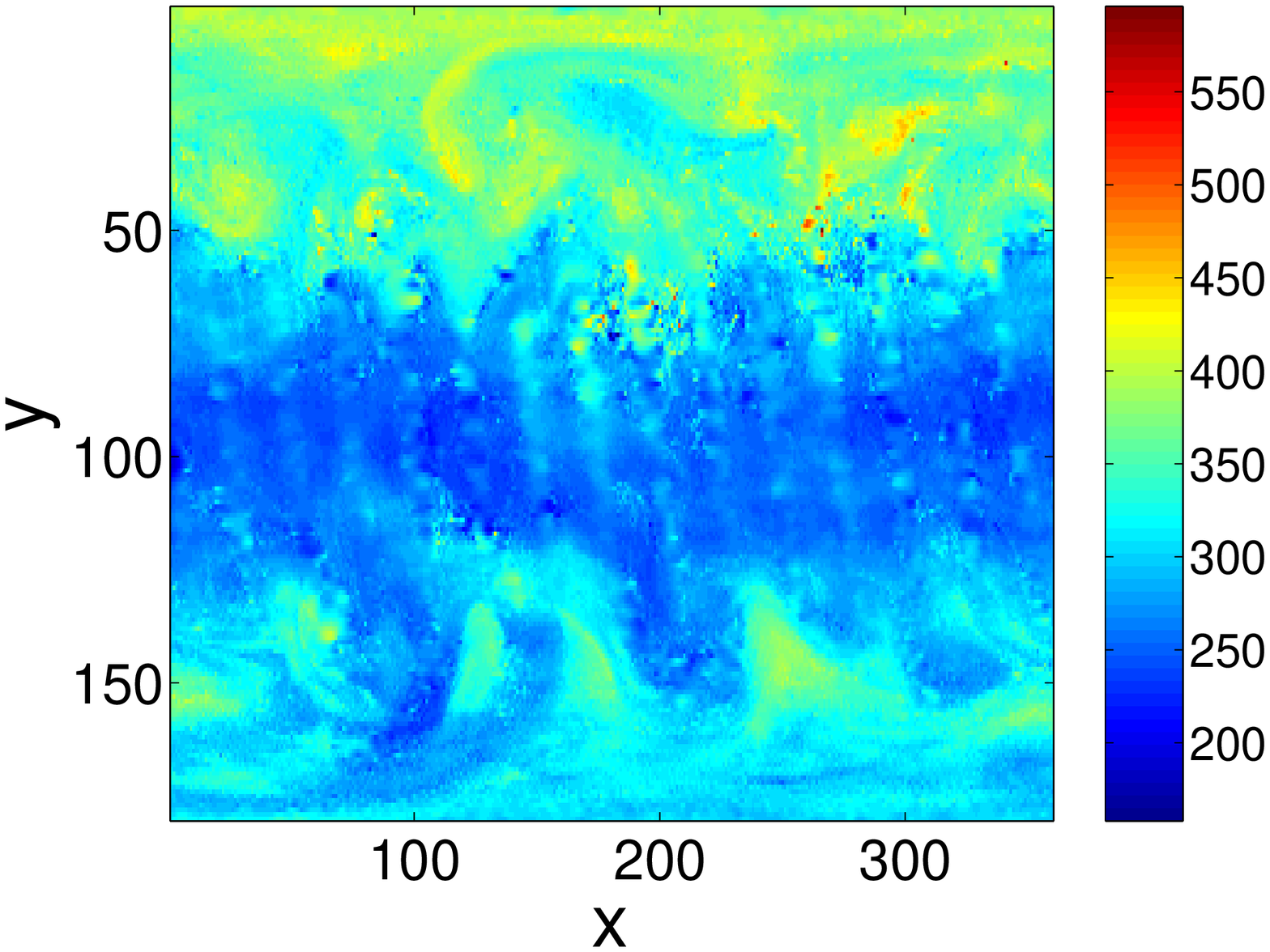}}
    \end{center}
%    \vspace{-10pt}
     \caption{Interpolation results for ozone data, using the nearest-neighbor (NN) (a),
     the inverse distance (ID) (b), and the biharmonic spline (BS) (c) methods.}
%     \vspace{-15pt}
    \label{fig:ozone_others}
\end{figure}

\begin{figure}[h!]
\begin{center}
    \includegraphics[scale=0.5]{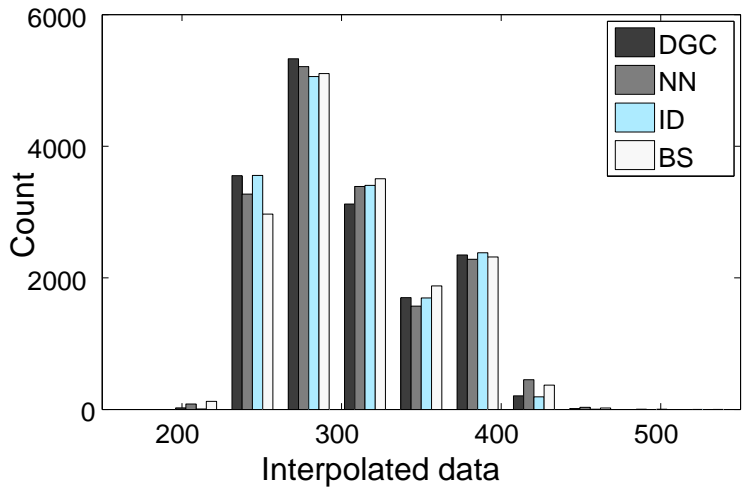}
\end{center}
   \caption{Histograms of the ozone values predicted by the respective methods: DGC, NN,
     ID, and BS.}
    \label{fig:ozone_hist}
\end{figure}

In the DGC reconstructed image, as shown in Fig.
\ref{fig:ozone_DGC_b}, some traces of the stripe pattern due to
overestimation in low-value areas (averaging effect) can still be
observed. However, this effect is somewhat less pronounced than in
other interpolation methods, presented in Fig.
\ref{fig:ozone_others}. Indeed, histograms of the predicted values,
see Fig. \ref{fig:ozone_hist}, show a larger proportion of DGC
predictions in bins below the sample average $\bar{z}=311.58$
compared to the other methods.

\section{Discussion and Conclusions}
\label{conclusion} We presented and investigated the DGC method for
the prediction of missing data on rectangular grids. DGC is based on
stochastic simulation conditioned by sample data with a global
objective function that accounts for anisotropic correlations. The
constraints involve normalized directional gradient and curvature
energies in specified directions. The simulation samples the
configuration subspace that leads to local minima of the objective
function.

For reliable application of DGC   sufficiently high sampling density
and number of data for the calculation of sample constraints are
desirable.  We evaluated the average numbers $n_{p}$ of
nearest-neighbor sampling-point pairs per direction and $n_{t}$  of
compact triplets of sampling points per direction. The first,
$n_{p}$, is equal to the number of terms involved in
$\overline{G}_{n}({\bf I}_{s})$, while the second, $n_{t}$, is the
number of terms in $\overline{C}_{n}({\bf I}_{s})$.  For uniform
random thinning $n_{p}$ and $n_{t}$ depend only on the degree of
thinning $p$ and the domain size $L$.  For $p=66\%$ we obtained
$(n_{p},n_{t}) = (280,92)$ for $L=50$ and $(n_{p},n_{t}) =
(7523,2542)$ for $L=256$ without significant differences between
different directions.  These values are sufficient for reliable
estimates of $\overline{G}_{n}({\bf I}_{s})$ and
$\overline{C}_{n}({\bf I}_{s})$. However, smaller grids or higher
thinning degrees can lead to insufficient sampling.

Regarding sensitivity of DGC to noise, we have run tests on
simulated random field realizations to which Gaussian white noise is
added. DGC seems more sensitive to noise than other interpolation
methods (e.g., BL, BC, BS), resulting in a larger increase of
cross-validation errors with increasing noise variance. This effect
is caused by the fact that methods like BL, BC, and BS perform some
smoothing of the noise by means of the weighted average over
extended neighbors. DGC on the other hand focuses on
 correlations over a small local neighborhood, which are sensitive to noise.
 Hence, in its current formulation DGC is more useful for smooth data
distributions, such as the ones studied herein. For noisy data,
improvements can be made by developing directional kernel-based
estimators for the square gradient and curvature in the spirit
of~\citep*{seldth08,dthsel09} or by incorporating an initial
filtering stage to reduce noise~\citep*{Brownrigg84,Yin96}.

DGC does not rely on assumptions about the probability distribution
of the data, it is reasonably efficient computationally, and it
requires very little user input (i.e., the number of simulation
runs, the number of class levels and the size of the maximum stencil
for initial state selection). Potential applications include filling
of data gaps in satellite images and restoration of damaged digital
records. For applications in the interpolation of data sampled on an
irregular grid, DGC needs to be extended to account for the lack of
grid structure. This can be accomplished using kernel functions with
adjustable bandwidth as shown in~\citep*{seldth08,dthsel09}.

DGC shares conceptual similarities with interpolation methods based
on splines, which are generated by minimizing an objective function
formed by the  linear combination of the square gradient and the
square curvature~\citep*{wessel09}. A special case of the
splines-based approach is the BS method used above for comparison
purposes. DGC does not require minimization of the square gradient
and curvature but requires matching the sample and entire-grid
values of these constraints. Splines-based methods require solving a
linear system involving the Green's function of the interpolation
operator; the numerical complexity of this calculation scales with
the third power of the system size. DGC does not require such a
costly operation, because the objective functional is defined using
local couplings. Finally, in contrast with splines interpolation,
DGC introduces a stochastic element by sampling the configuration
subspace of local minima of the DGC objective function. On the other
hand, splines-based methods can handle irregularly spaced samples,
while DGC is currently restricted to grid data.

\section*{Acknowledgments}
This work is funded by the European Commission,  under the 6th
Framework Programme, Contract N. 033811, DG INFSO, action Line
IST-2005-2.5.12 ICT for Environmental Risk Management. The views
expressed herein are those of the authors and not necessarily those
of the European Commission.

Ozone layer thickness data used in this paper were  produced with
the Giovanni online data system, developed and maintained by the
NASA Goddard Earth Sciences (GES) Data and Information Services
Center (DISC).

\section*{References}
%\bibliography{library}

\end{document}